\documentclass[acmtog, nonacm]{acmart}

\makeatletter
\def\@ACM@checkaffil{%
    \if@ACM@instpresent\else
    \ClassWarningNoLine{\@classname}{No institution present for an affiliation}%
    \fi
    \if@ACM@citypresent\else
    \ClassWarningNoLine{\@classname}{No city present for an affiliation}%
    \fi
    \if@ACM@countrypresent\else
        \ClassWarningNoLine{\@classname}{No country present for an affiliation}%
    \fi
}
\makeatother

\usepackage{booktabs} %
\usepackage{multirow}
\usepackage{makecell}
\usepackage{framed}
\usepackage{color}

\usepackage[linesnumbered,ruled]{algorithm2e} %

\usepackage{tikz}
\usetikzlibrary{fit,calc,backgrounds}

\IncMargin{1em}

\SetAlFnt{\small}
\SetAlCapFnt{\small}
\SetAlCapNameFnt{\small}
\SetAlCapHSkip{0pt}

\acmJournal{TOG}

\definecolor{myorange}{RGB}{251, 207, 157}
\definecolor{mypink}{RGB}  {250, 179, 195}
\definecolor{myyellow}{RGB}{255, 242, 204}

\def\best#1{\colorbox{mypink}{#1}}
\def\second#1{\colorbox{myorange}{#1}}
\def\third#1{\colorbox{myyellow}{#1}}

\begin{document}

\title{NEPHELE: A Neural Platform for Highly Realistic Cloud Radiance Rendering}

\author{Haimin Luo}

\author{Siyuan Zhang}

\author{Fuqiang Zhao}

\author{Haotian Jing}

\author{Penghao Wang}

\author{Zhenxiao Yu}

\author{Dongxue Yan}

\author{Junran Ding}

\author{Boyuan Zhang}

\author{Qiang Hu}

\author{Shu Yin}

\author{Lan Xu}

\author{Jingyi Yu}
\affiliation{
\institution{ShanghaiTech University}
\country{China}
}

\renewcommand\shortauthors{Luo, H. et al}

\begin{abstract} 
We have recently seen tremendous progress in neural rendering (NR) advances, i.e., NeRF, for photo-real free-view synthesis. Yet, as a local technique based on a single computer/GPU, even the best-engineered Instant-NGP or i-NGP cannot reach real-time performance when rendering at a high resolution, and often  requires huge local computing resources. 
In this paper, we resort to cloud rendering and present NEPHELE, a neural platform for highly realistic cloud radiance rendering. In stark contrast with existing NR approaches, our NEPHELE allows for more powerful rendering capabilities by combining multiple remote GPUs, and facilitates collaboration by allowing multiple people
to view the same NeRF scene simultaneously.
Such a combination of NeRF and cloud rendering naturally requires a lightweight, real-time neural renderer with flexible scalability. To this end, analogous to i-NGP, we introduce i-NOLF to employ opacity light fields for ultra-fast neural radiance rendering in a one-query-per-ray manner. We further resemble the Lumigraph with geometry proxies for fast ray querying, and subsequently employ a small MLP to model the local opacity lumishperes for high-quality rendering. We also adopt Perfect Spatial Hashing in i-NOLF to replace the brute-force multi-hashing in the original i-NGP, so as to enhance cache coherence. As a result, our i-NOLF achieves an order of magnitude performance gain in terms of efficiency than i-NGP, especially for the multi-user multi-viewpoint setting under cloud rendering scenarios. 
We further tailor a task scheduler accompanied by our i-NOLF representation, with a ray-level scheduling design to maintain the resiliency of rendering jobs.
We also demonstrate the advance of our methodological design through a comprehensive cloud platform, consisting of a series of cooperated modules, i.e., render farms, task assigner, frame composer, and detailed streaming strategies.
Using such a cloud platform compatible with neural rendering, we further showcase the capabilities of our cloud radiance rendering through a series of applications, ranging from cloud VR/AR rendering, to sharing NeRF assets between multiple users and allowing NeRF assets to freely assemble into a new scene.

\end{abstract}


\begin{teaserfigure}
  \centering
\includegraphics[width=1.0\linewidth]{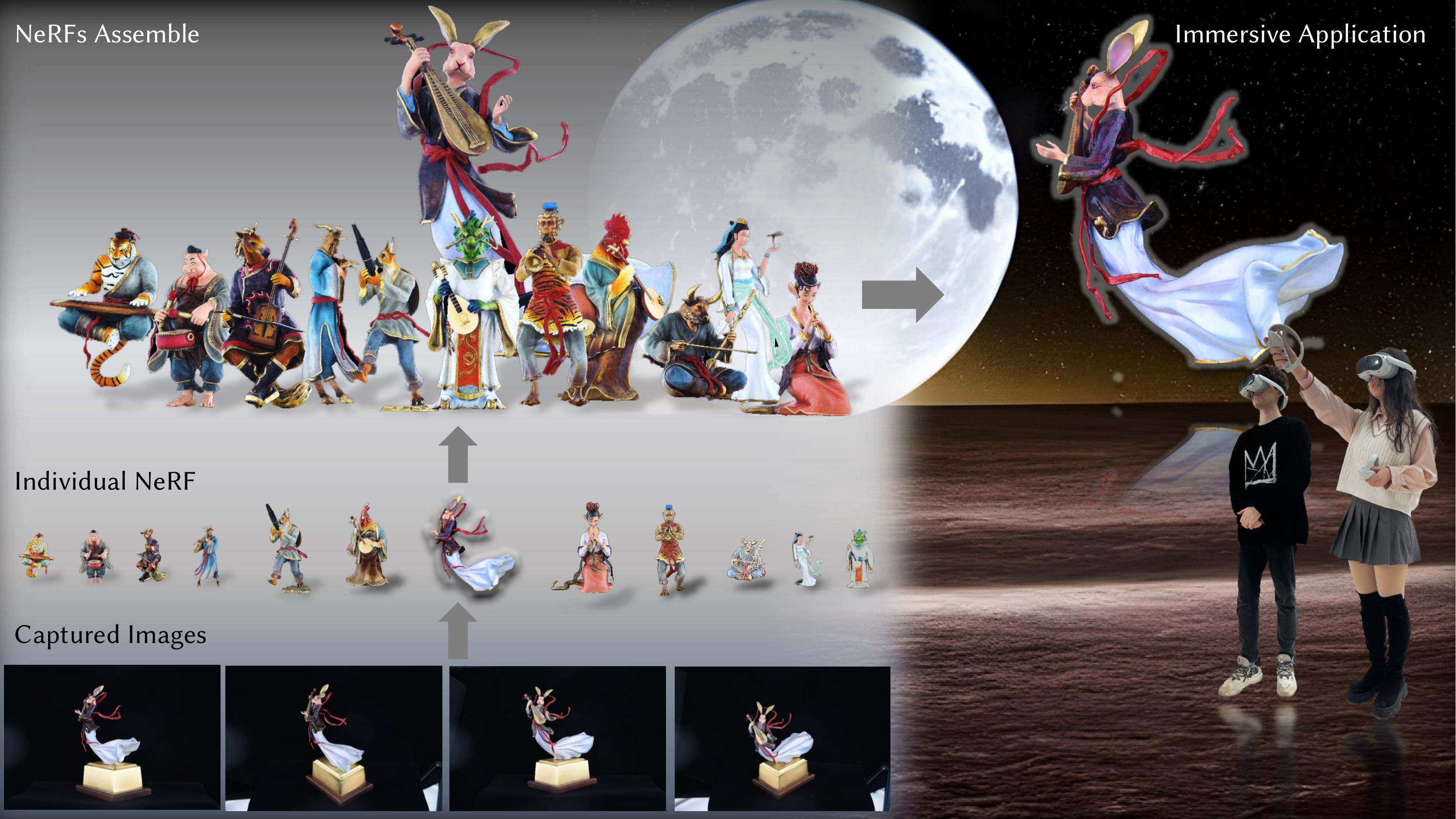}
  \caption{\textbf{Demonstration of our NEPHELE platform}. We present NEPHELE, a neural platform for highly realistic cloud radiance rendering. Our approach represents the objects, e.g., the \textbf{Chinese zodiac}, as i-NOLFs for ultra-fast neural radiance rendering. With a tailored task scheduler, NEPHELE allows for utilizing the power of multiple GPUs to render a complex scene composed of many neural assets which cannot be achieved by a single GPU. NEPHELE further distributes the photo-realistic radiance renderings to the clients so that everyone can easily access the shareable neural assets on the cloud via various devices such as VR headsets and mobiles.
  }
  \label{fig:teaser}
\end{teaserfigure}

\maketitle

\section{INTRODUCTION}
As neural rendering is rapidly gaining momentum to replace conventional physical-based approaches, so are the demands for more accessible and more affordable access for practical uses. By employing tailored deep networks to simultaneously model scene geometry and appearance in terms of rays, radiance field based approaches such as NeRF map rendering to ray querying through the network. They can reproduce complex real-world objects and even large environments with high photorealism. Yet, as a local technique based on a single computer/GPU, even the best-engineered instance-NGP cannot reach real-time performance when rendering at a high resolution: on a single NeRF object, i-NGP can barely get 10 fps at 4K resolution even on the latest NV GeForce 4090.

In this paper, we resort to cloud rendering and present NEPHELE, a NEural Platform for Highly rEalistic cLoud radiance rEndering. In a nutshell, NEPHELE uses remote servers rather than the processing power of a single local computer to not only allow for more powerful rendering capabilities by combining multiple GPUs but also realize a range of unprecedented effects. For example, NEPHELE enables the rendering of complex scenes composed of many NeRF objects that would be too demanding for a single computer. More importantly, it facilitates collaboration by allowing multiple people to view the same NeRF scene simultaneously. Further, by enabling the rendering process to run in the background, NEPHELE frees up the user's computational for other tasks, e.g., to support a variety of virtual and augmented reality experiences. 

Traditional cloud rendering is largely based on a rasterization-based rendering pipeline. They essentially split the rendering process into smaller tasks, which are then distributed across multiple servers. Specifically, the 3D scene and assets are first divided into smaller, manageable subsets where each subset can be rendered in real-time at individual servers. The rendered results are then sent back to the cloud platform, where the final rendered images are assembled. However, direct deployment of classical cloud solutions to neural radiance field rendering faces multiple challenges. First, the smallest "subset" in NeRF is a neural network that commonly represents the complete object, and one would have to allocate one graphics card for each network. Even so, real-time rendering either sacrifices resolution~\cite{chen2022mobilenerf, Artemis} or requires ultra-high storage~\cite{yu2021plenoctrees, yu_and_fridovichkeil2021plenoxels,chen2022mobilenerf}. Therefore, it is essential to have a lightweight, real-time neural renderer to maintain flexibility and scalability. Second, for an immersive experience, multiple users should be able to view the same NeRF object simultaneously but from different perspectives. Hash-encoding, designed to support a single user in i-NGP, can lead to a high miss rate and degrades the rendering performance. Finally, schedulers should assign multiple rendering jobs to best-fit graphics cards for higher hardware utilization, better rendering performance, and good data coherence by viewing the same NeRF object. But existing works focus on scheduling policies in the unit of jobs, which introduces context switch overhead and poor utilization of graphics cards. Therefore a more tailored scheduler would maintain the resiliency of rendering jobs.

In NEPHELE, we first develop an ultra-fast neural radiance renderer called Instant Neural Opacity Light Fields or i-NOLF. i-NOLF conducts a one-query-per-ray scheme by marrying the opacity light fields\cite{OLF} and i-NGP to significantly accelerate hit point estimation and, subsequently, the overall rendering. It resembles the Lumigraph~\cite{buehler2001unstructured} that uses geometry proxies for fast ray querying and employs an additional small MLP to model the local opacity lumishperes for supporting view-dependent features such as translucency and specularity.  By using an efficient data structure to cache the view-independent properties, i-NOLF achieves over half an order of magnitude acceleration to i-NPG and can render at a 4K resolution over 30 fps. To further overcome the limit of multi-resolution hash encoding~\cite{NGP}, we show significant cache traffic under the multi-user multi-viewpoint setting is caused by accessing short features evenly distributed across different levels of the hash table. Instead, we adopt the Perfect Spatial Hashing ~\cite{10.1145/1141911.1141926} to reorganize the memory distribution to enhance cache coherence and develop a single-resolution encoding scheme to reduce the cache traffic. Putting two schemes together, i-NOLF is an order of magnitude faster than i-NGP under the multi-user setting.

At the system level, an efficient multi-user cloud renderer relies on intelligent task distribution/scheduling and result gathering to optimize the response time while improving the utilization of GPUs in the render farm.
We propose a ray-level rendering mechanism that treats multi-user rendering tasks as an indiscriminate range of rays with depths. In such a way, our scheduler can maintain a higher efficiency of GPU by scheduling multiple rendering tasks by ray-level metrics. Second, we present a ray-level scheduling design that contains a centralized scheduler.
The master node merge tasks with commonness to make use of cache
and compose the results with correct ordering.

NEPHELE aims to distribute the power of radiance rendering via the cloud, analogous to sunlight radiating through the cloud over the earth. One of the target applications in neural rendering is providing users with an immersive experience under either augmented or virtual reality settings. To seamlessly connect NEPHELE with commodity solutions such as the Apple ARKit~\cite{arkit} and WebXR device API~\cite{webxr}, we adopt a streaming-based solution where computationally intensive workloads are conducted on cloud servers, and their results are sent to the client as a video stream. To maintain low latency and interactivity, we employ Web Real-Time Communication (WebRTC) \cite{webrtc} where servers and clients can directly communicate with each other. %
For our cloud AR application, the mobile device uses the ARKit to continuously deliver user interaction data, e.g., transformation matrix, intrinsic matrix, and detected plane information, to the cloud server. The cloud server renders the content with a pure green background based on the interaction data and compresses the rendered frames into a video stream transmitted to the mobile device through WebRTC. The mobile device decodes the video frames and performs background matting with chroma keying to make a specific scope of the decoded frames transparent. Finally, we overlay the frames onto the camera screen for an immersive experience that blends the physical and digital worlds. 
For our cloud VR application, we initialize the cloud task with device resolution and camera projection matrix, then continuously receive current views and poses from the device. The cloud task renders a scene for each view and returns the combined frame and depth information through WebRTC. The web application creates texture, material, and mesh based on the returned data and renders the scene with the mesh. Finally, we showcase the capabilities of NEPHELE to distribute the power of radiance rendering via the cloud for multi-user scenarios, ranging from experiencing and sharing NeRF assets between multiple users to allowing NeRF assets to assemble into a new scene freely.

To summarize, our main contributions include the following:
\begin{itemize} 
\setlength\itemsep{0em}
    \item We propose NEPHELE, a cloud rendering platform, which is  designed to distribute the power of radiance rendering via the cloud, allowing for more powerful rendering capabilities.

    \item We present an ultra-fast neural radiance renderer, i-NOLF, to employ the opacity light fields and Perfect Spatial Hashing into the i-NGP. It significantly outperforms i-NGP, especially for the multi-user multi-viewpoint setting.
	
    \item We tailor a task scheduler accompanied by our i-NOLF representation with a ray-level scheduling design to maintain the resiliency of rendering jobs.

    \item We showcase the capabilities of cloud radiance rendering through a series of applications, i.e., cloud VR/AR rendering or sharing and assembling various NeRF assets.
    
\end{itemize}

\section{Related Work} \label{sec:related}

\paragraph{Neural radiance rendering}
The recent progress of Neural Radiance Fields (NeRF)~\cite{mildenhall2021nerf} technique, and its extensions~\cite{convnerf,verbin2022refnerf,chen2021mvsnerf,wang2021ibrnet,barron2021mip} bring huge potential for photo-realistic novel view synthesis. 
Its dynamic extension\cite{zhang2021editable, zhao2022humannerf,pumarola2020d,tretschk2021nonrigid,liu2021neural,peng2021neural} focus on dynamic scenes with NeRF volume rendering techniques.
Some works \cite{nerv2021,zhang2021nerfactor,pan2021shadegan,boss2021nerd} extend NeRF with lighting decomposition, enabling NeRF with relighting effects.
Other following 
works~\cite{lombardi2021mixture,yu2021plenoctrees,sun2022direct, wang2022fourier,yu_and_fridovichkeil2021plenoxels,NGP,chen2022tensorf} have improved the training and rendering speed of NeRF witch make NeRF representation more efficient and practical.
For example, PlenOctrees~\cite{yu2021plenoctrees} converts NeRF continues representation into a discrete octree representation. 
It achieves $>$ 3000$\times$ rendering speedup using this efficient GPU octree implementation without any MLP evaluations.
i-NGP~\cite{NGP} further improves training speed by combining multi-resolution hashing encoding and tiny MLPs.
It helps i-NGP~\cite{NGP} reduce training time to 5 seconds while keeping $>$ 3000$\times$ rending speedup compared to the original NeRF.
However, these methods still require high-end GPU devices to support computation-intensive training and rendering process.
MobileNeRF~\cite{chen2022mobilenerf} proposed a new NeRF representation based on a set of polygons with textures that consist of binary opacities and feature vectors.
It can render novel images efficiently on mobile devices such as Phones and light Laptops.
However, it still relies upon a high-end GPU to optimize such a textured polygons-based NeRF representation.
In addition, this proposed textured polygons-based representation also lost the view-dependent effectiveness.
Most existing related approaches focus on a single object or scene.
Fewer methods~\cite{zhang2021editable,niemeyer2021giraffe,guo2020object} enable NeRF representation with compositionality.
To this end, we aim to build a NeRF cloud rendering platform (NEPHELE) that eliminates users' dependence on high-end devices and makes NeRF techniques more available.
The proposed NEPHELE keeps real-time NeRF rendering in VR/AR applications and hologram displays such as looking glass.
Besides, NEPHELE also entrusts compositionality to each NeRF object on this cloud platform.

\begin{figure*}
	\includegraphics{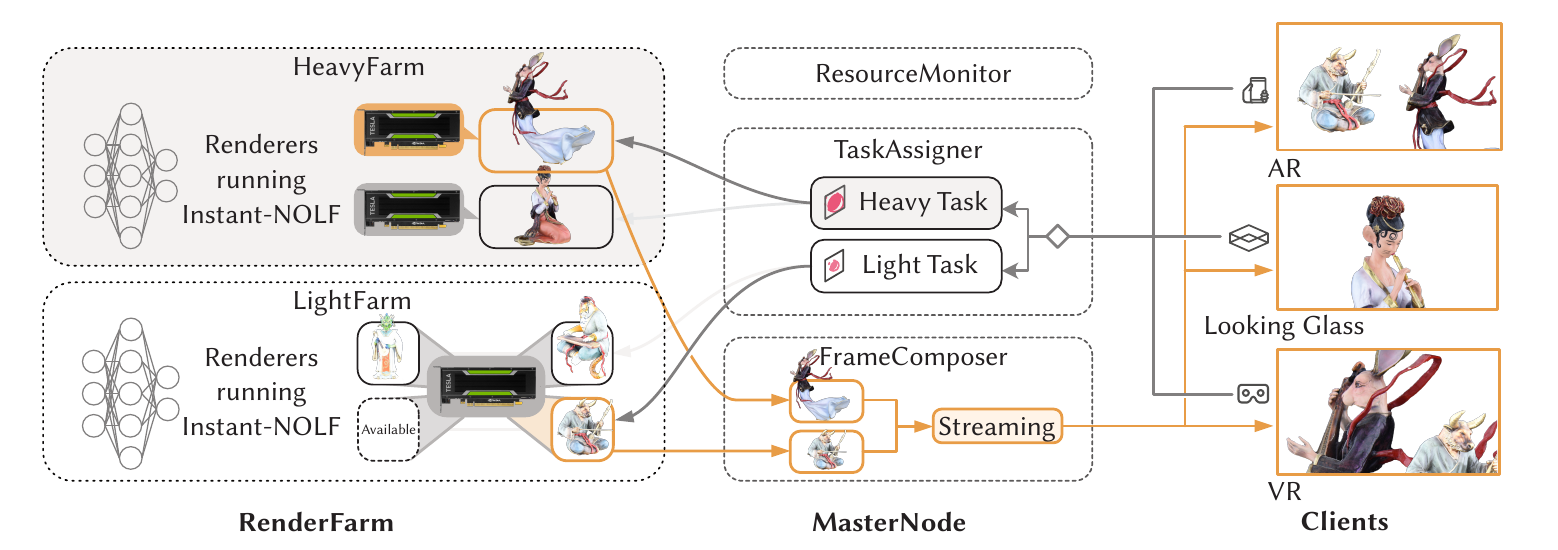}
	\caption{\textbf{Overview of our NEPHELE system.} 
 A flowchart depicting the process of rendering a scene in a cloud platform, where the user uses a client to connect to the \textbf{MasterNode}, which creates a session and establishes a WebRTC connection. The user makes a request by submitting their scene and viewing objects. The \textbf{TaskAssigner} then categorizes the tasks as light or heavy based on the approximate number of rays that hit the density grid. Light tasks are routed to a light farm for rendering, while heavy tasks are routed to a heavy farm. The \textbf{RenderFarm} renders the tasks into frames, which the \textbf{FrameComposer} then assembles to match the user's request. Finally, the \textbf{FrameComposer} sends the finished frame to the client to be viewed.
 \textit{
 The solid black rectangle represents valid GPU resources. The orange rectangle indicates that a GPU is monopolized for a heavy task (HeavyFarm) or part of a GPU is occupied for a light task. The dashed rectangle represents the GPU resources available for the newly assigned task.}
 }
	\label{fig:pipeline}
\end{figure*}

\paragraph{Task scheduling for GPU clusters}

Many studies have targeted GPU cluster schedulers. SLAQ \cite{SLAQ} and Optimus \cite{Optimus} schedule distributed jobs to minimize the average computing time by estimating the future steps of each job. Gandiva \cite{Gandiva} and Themis \cite{Themis} move one step further to consider completion time fairness with time-slicing mechanisms on GPU resources. Some other schedulers spend efforts on developing placement policies based on the priority of jobs for better locality performance. For example, Philly \cite{Philly} attempts to fine-tune the scheduling policy to alleviate average job completion time based on the trace analysis from production clusters. Tiresias \cite{Tiresias} uses attained service time of jobs to form priority queues for scheduling. ONES \cite{ONES} determines the batch size of each job and proposes an online evolutionary scheduler for elastic workloads. Some early works present schedulers based on greedy policies to maximize the overall system utilization, such as OASiS \cite{OASiS}. All these works focus on scheduling policies in the unit of jobs due to the diversity of applications. Besides, placing an entire job on dedicated GPU resources retains good data consistency and protection via isolation. But it introduces context switch overhead to the GPU clusters and leads to lower utilization of GPU resources. We noticed a difference between rendering jobs and general GPU-based applications: a certain amount of data is shareable amongst multiple rendering jobs, especially viewing the same objects from different angles. Based on this observation, Our proposed NEPHELE system offers a placement policy that treats rendering jobs in groups of rays. With the help of the finer rendering granularity, NEPHELE schedules jobs considering multiple factors (i.e., fairness, responsiveness, job completion time, and total utilization).  

\paragraph{Streaming}
Cloud-based rendering technology is widely used for render farms nowadays, with the potential to reduce hardware requirements on the client side and free users from unlimited hardware upgrades. 
However, this potentially revolutionizing computing paradigm could become a huge failure without an appropriate streaming solution designed for cloud-based rendering applications. 
Interaction latency is one of the main issues for cloud-based rendering. 
The user interaction is sent to the cloud, where a scene will be rendered according to the user interaction. 
The rendered frames on the cloud should be encoded and  delivered to the client as quickly as possible.
Apple HTTP Live Streaming (HLS) \cite{hls} and MPEG Dynamic Adaptive Streaming over
HTTP (DASH) \cite{DASH} are two of the most popular HTTP adaptive streaming protocols.
They work by splitting the video into smaller files, called segments which can be instantly downloaded by the client. 
This results in about seconds of glass-to-glass delay. Therefore, HLS and DASH are not appropriate for cloud-based rendering.

 To support these low-latency and interactive use cases, Real-Time Communication (RTC) solutions are generally used.
Web Real-Time Communication (WebRTC) is an open-source project by Google, which maintains a simple peer-to-peer architecture to achieve efficient communication \cite{webrtc}.
 WebRTC has low latency, while it involves a complete set of protocol designs and QoS guarantee mechanisms. With the development of WebRTC technology, many related applications such as remote XR \cite{xr1,xr2}, telemedical \cite{app10010369}, and cloud gaming \cite{network1030015}  have emerged because of its fast and stable transmission speed. 
 Our work also leverages the characteristics of WebRTC, which makes it possible to experience low latency and high-quality cloud rendering services.

\section{Overview: NEPHELE Architecture} \label{sec:overview}
\begin{figure}
    \includegraphics{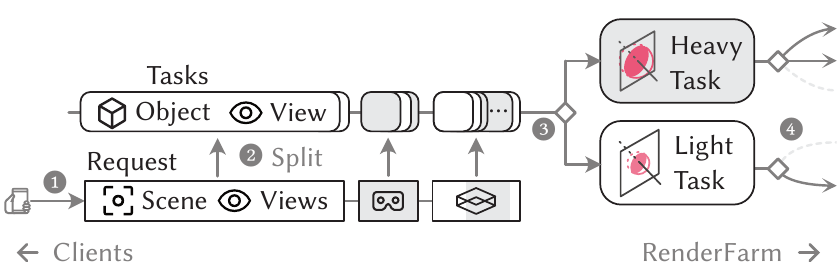}
    \caption{\textbf{Function of TaskAssigner.} It receives requests from clients, including information about the scene and the desired views. The requests are then broken down into individual tasks containing an object with angles. Tasks are classified as either light or heavy based on the estimated number of rays that will hit the density grid. Light tasks are sent to a dedicated light farm for rendering, while heavy tasks are routed to a heavy farm for processing.}
    \label{fig:task-assigner}
\end{figure}%
\subsection{Design Goals}
First, NEPHELE should support simultaneous multi-user rendering tasks, where multiple users generate random access to the GPU cluster for multi-angle viewings. Second, NEPHELE should treat independent rendering tasks as a group of indiscriminate rays with depth, where the boundary of tasks can be neglected for better data re-usability. Third, NEPHELE should schedule tasks in the unit of a group of rays to GPUs for better utilization and response time.

\begin{figure*}[t]
  \includegraphics[width=1.0\linewidth]{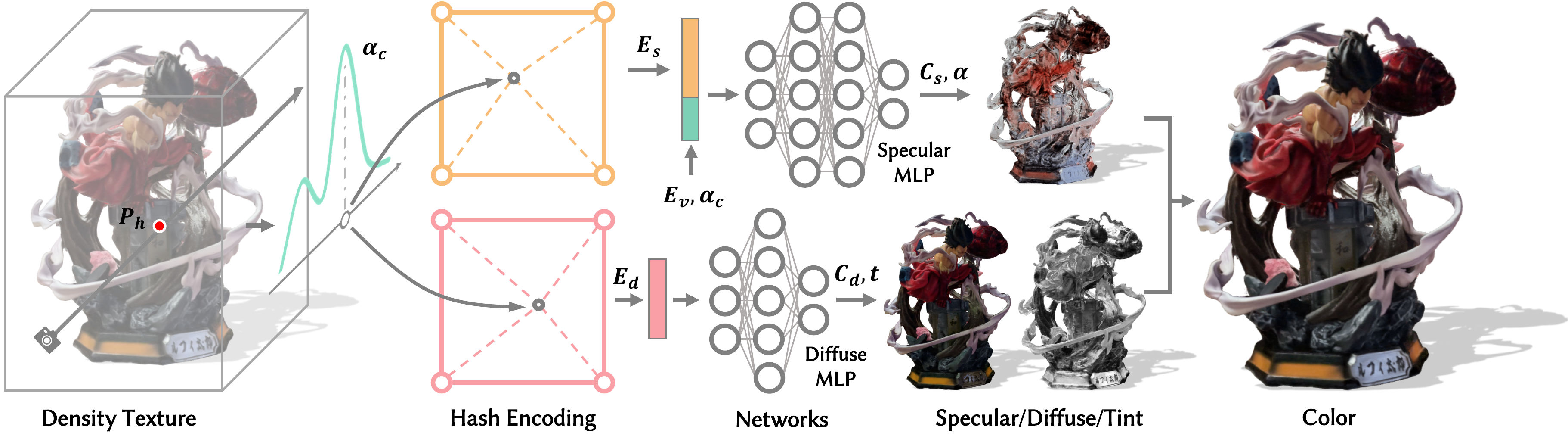}
  \caption{\textbf{The illustration of our i-NOLF to render the Chinese mythological character \textit{Nezha}.} For a given ray, we first find the hit point coordinate $P_h$ using ray-marching on the cached density texture and accumulate a coarse opacity $\alpha_c$ of the ray. The resulting coordinate is then encoded to a feature vector $E_s$ by a specular hash encoder. We concatenate $E_s$ with a view direction encoding $E_v$ and $\alpha_c$. The specular MLP then maps the resulting feature to specular color $C_s$ and opacity $\alpha$. Besides, we encode $P_h$ to $E_d$ with another diffuse hash encoder and then adopt a diffuse MLP to predict diffuse color $C_d$ and specular tint $t$. The diffuse and specular colors are then combined into the final outgoing radiance.}
  \label{fig:nolf}
\end{figure*}

\subsection{NEPHELE Architecture}
Fig. \ref{fig:pipeline} illustrates the architecture of NEPHELE. NEPHELE consists of three major components: RenderFarm performs render tasks in the unit of a group of rays, MasterNodes monitor the status of rendering tasks, assign tasks to target GPUs in the RenderFarm, and compose rendered results with correct ordering, and Clients represent diverse devices to submit rendering jobs and display results (e.g., PC, mobile phones, AR devices, VR devices, etc.), 

\textbf{RenderFarm} emphasizes that users can \textbf{share} the \textbf{same process} to render the same NeRF model, achieving the highest hardware utilization, using similar rendering results between multiple users to reduce unnecessarily repeated computation, and reducing per-user computation costs, in contrast to traditional cloud game rendering or cloud render farms where each user's rendering task is isolated on a different process or even on completely isolated hardware.

\textbf{MasterNode} is in charge of monitoring the status of RenderFarm, assigning rendering tasks to the target GPUs in RenderFarm, and composing the ray-based rendering results with correct ordering for Clients. Typically, ResourceMonitor surveils the utilization of GPUs in RenderFarm and informs TaskAssigner to distribute rendering tasks to target GPUs. 

TaskAssigner in MasterNode allows a single GPU to run numerous low-precision NeRF instances and multiple GPUs to run a small number of high-precision ones. TasksAssigner first transforms the rendering job into a group of rays with angles and categorizes rays into Heavy and Light tasks based on their vision coverage. Heavy indicates that a large number of rays hit the object, while Light implies that only a few rays hit it. A Heavy task could be a user staring at a small entity at close quarters or looking at a large one, and a Light one could be a user overlooking a large object or observing a small one. Scheduling details are in Section \ref{sec:scheduling}.

\textbf{Clients} are the multiple rendering jobs submitted by diverse users. The jobs can be a single visitor wandering relics in a museum with AR (a.k.a. single-user-multiple-object) or HMD devices or a group of tourists examining the same artwork in a gallery (a.k.a. multiple-user-single-object). Existing rendering techniques usually monopolize one or more GPUs for a single job due to the overhead of initialization and context switch of NeRF models. It is challenging to render numerous objects in a scene for a single user or multiple users because it requires unaffordable GPU resources. Although we can place and render each model independently on GPUs and composite together for a large scene, the overhead of context switching and communication amongst GPUs is too expensive to provide an affordable and prompt NeRF rendering experience. 

Besides, when a single user attempts to view different objects in the same breath or multiple viewers for the same object concurrently, the multiple rendering tasks generated lead to simultaneous random access to GPUs. It causes an unacceptable response time because the cloud scheduler queues all rendering tasks for GPUs. Newly submitted rendering tasks have to wait until sufficient GPU resources are released. We notice a decent number of rays are sharable in different viewing tasks in both the cases of single-user-multiple-object and multiple-user-single-object. If multiple rendering jobs can share the data representing similar viewing angles, we can reduce redundant GPU rendering labor and take advantage of the GPU cluster for more jobs. It requires RenderFarm to execute rendering jobs with a finer-grain rendering algorithm so that MasterNode can schedule the tasks in the unit of a group of rays instead of tasks.

\section{Accelerating i-NGP} \label{sec:accelerating}

As the core of the RenderFarm of NEPHELE, a fast neural renderer is essential to render realistic images in as short a duration as possible to meet the user's requirements. 
We set out to accelerate the recent real-time neural renderer i-NGP. We first improve the memory request efficiency of i-NGP by developing a novel perfect spatial hashing encoding. We further present an ultra-fast renderer by marrying the opacity light fields with i-NGP to conduct one-query-per-ray rendering to avoid redundant computational overhead.

\begin{figure}[t]
  \includegraphics[width=\linewidth]{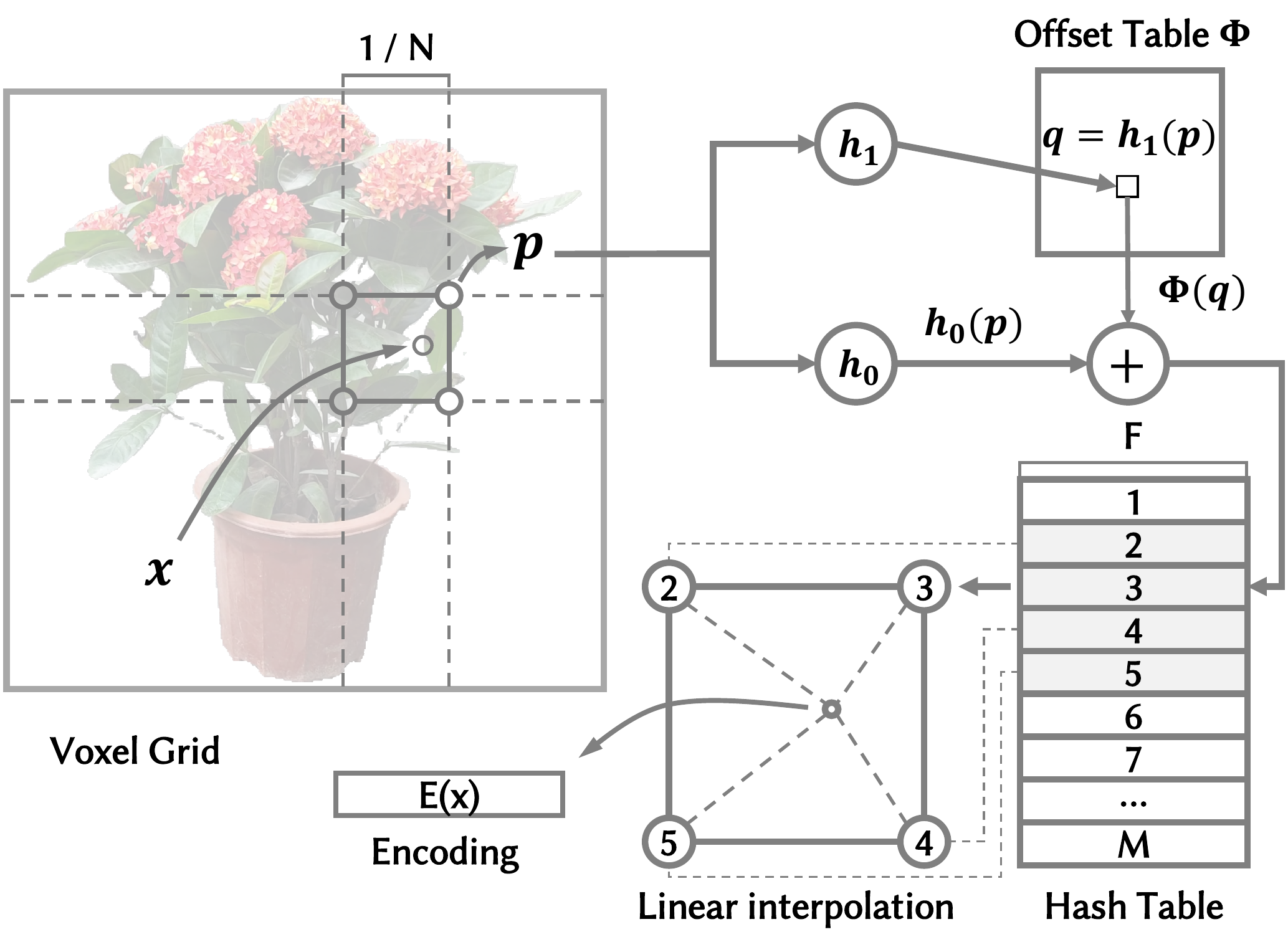}
  \caption{\textbf{Illustration of perfect spatial hashing encoding in 2D}. For a given input coordinate $x$, we find the corners of the voxel where $x$
 is located and hash their integer coordinates $p$ with two hashing functions, $h_0$ and $h_1$, to base indices to hash table and indices $q$ to offset table $\mathbf{\Phi}$ respectively. We look up the corresponding F-dimensional feature vectors with the resulting corner indices $h_0(p) + \mathbf{\Phi}(q)$ from the hash table and generate the final encoding $E(x)$ by tri-linear interpolation. Note that spatially adjacent points are mapped to adjacent slots in the hash table without hashing collisions to maximize cache coherence.}
  \label{fig:psh}
\end{figure}

\subsection{Perfect Spatial Hashing Encoding}

\paragraph{Limitations on multiresolution hash encoding.} 
Recall that in i-NGP~\cite{NGP}, multiresolution hash encoding adopts uniform grid encoding at coarse resolution, while hash encoding is used at fine resolution.
The short feature vectors from different resolutions are concatenated to the final feature vector, yielding a multiresolution feature vector with a manageable memory footprint.

However, utilizing GPU to implement such encoding is inefficient.
According to the property of the hash function, memory access to the hash table is distributed evenly across the entire hash table. Cache thrashing will therefore occur if the hash table can not fit in the cache, resulting in significant memory to cache traffic.
Moreover, it will request memory independently at each level but only use a small fraction of the cache line returned, wasting memory bandwidth.

\paragraph{Perfect Spatial Hashing. } Thus, to create cache coherence and to better utilize memory bandwidth, we present a novel single-resolution perfect spatial hashing encoding, as shown in Fig.~\ref{fig:psh}.
Perfect spatial hashing (PSH) refers to the hash function precomputed on static data, e.g., occupancy field, to have zero hash collisions.
We adopt hash function of form ~\cite{10.1145/1141911.1141926}
\begin{equation}
    h(\mathbf{p}) = h_0(\mathbf{p}) + \Phi(h_1(\mathbf{p}))
\end{equation}
where $\Phi$ is a lookup table for offset constructed by a heuristic approach inspired by \cite{Perfecthashing} to help minimize hash collisions.
With the offset table $\Phi$, PSH tends to find one that maximizes the case where spatially adjacent points are mapped to adjacent slots in the hash table, thus creating chances to exploit cache coherence.

Similar to hash encoding, for any position $\mathbf{x}$, we scale it by grid resolution $N$ and round to find the vertex coordinate $\mathbf{p} = \lfloor N*\mathbf{x} \rfloor$.
Then, using PSH, the associated feature vector is extracted for each vertex of the grid.
Lastly, the feature vectors are tri-linearly interpolated to get the result feature vector.
To reduce the number of memory requests, we only use single-resolution feature vectors.
So that for each sample point, we only issue one memory request and will consume half of the cache line returned.
\paragraph{Implementation for i-NGP}
We replace the multiresolution hash encoding with the proposed PSH encoding to evaluate the performance. Specifically, we perform the Shape-from-Silhouette (SfS) algorithm to obtain a coarse occupancy field from which the offset table $\Phi$ is computed. We train the models on the scene "ficus" with different PSH encoding resolutions and the original i-NGP. As shown in Table.~\ref{tab:PSH}, the finer resolution results in better rendering quality, and overall the performance is comparable with the hashgrid. However, as for the memory request efficiency, i.e., L2 cache hit rate, the PSH encoding significantly outperforms the hashgrid while testing with heavy concurrent random rendering requests. The rendering speed is 20\% faster at 4k resolution, which meets the requirements of the cloud rendering scenarios.

\subsection{Instant Neural Opacity Light Fields}

The main drawback of NeRF (including i-NGP~\cite{NGP}) is that it requires hundreds of neural network forward passes for rendering a pixel, even with an early-stop acceleration strategy. It results in both redundant GPU memory I/O operations and computational overhead in the ray-marching process, which is the major bottleneck to further speed up the rendering process in cloud scenarios. 
A straightforward way~\cite{R2L} is to represent the scenes as neural light fields (NeLF) in pursuit of one-query-per-pixel rendering. Such approaches usually require a large global fully connected neural network to encode the entire light field without geometry, which goes against the key design philosophy of i-NGP of using a smaller neural network under the premise of efficient spatial hash encoding. Besides, the hash grid tends to effectively encode local spatial properties but cannot approximate the global light field functions. 
To this end, we set out to represent the scenes as local neural opacity light fields to utilize the facilities of spatial hashing encoding and light field at the same time.

\setlength{\tabcolsep}{6pt}
\begin{table}[t]
\centering
\caption{\textbf{Performance of PSH encoding in different resolution}. We evaluate the PSH encoding by replacing the original hashgrid encoding of Instan-NGP and train the models on the scene "ficus". Besides the rendering quality, we evaluate the memory efficiency by L2 cache hit rate and rendering speed on rendering random 4K resolution images. We also report the number of parameters of the encoders.}
\resizebox{1\linewidth}{!}{
\begin{tabular}{lcccccc}
\toprule
Method & PSNR$\uparrow$ & SSIM$\uparrow$ & L2$\uparrow$ & size(MB) $\downarrow$ & time(ms) $\downarrow$ & FPS$\uparrow$\\ \hline
Hashgrid        & 32.90 & 0.892 & 72.82 & 12.20 & 69.54 & 14.38\\
\hline
PSH-128         & 31.61 & 0.960 & 94.18 & 1.37 & 56.09 & 17.83\\ 
PSH-256         & 32.31 & 0.967 & 94.33 & 7.63 & 56.30 & 17.76\\ 
PSH-512         & 32.40 & 0.968 & 94.23 & 51.25 & 58.55 & 17.08\\ 
\bottomrule 
\end{tabular}
}
\label{tab:PSH}
\end{table}

\paragraph{Opacity Light Field.} 
Early image-based rendering techniques essentially aim to model the plenoptic function by a surface light field~\cite{SLF, DSLF}, which encodes the outgoing radiance of a point located at a known mesh surface as a lumisphere. 
To get rid of the limitations of the complexity of the object's geometry, opacity light field~\cite{OLF} combines surface light field with view-dependent shape using opacity hulls~\cite{IBOH}. Opacity hulls use view-dependent opacities for surface points on proxy geometry to model elaborate geometrical details, e.g., feathers. Roughly speaking, an opacity light field can be represented as a function:
\begin{equation}
    L: M\times S^{2}\rightarrow RGBA
\end{equation}
where $M$ is a surface mesh, and $S^2$ denotes the sphere of unit vectors. 
In contrast to using an explicit mesh and view-dependent textures, we represent a continuous opacity light field as a 5D function which takes a 3D location $\mathbf{p} = (x, y, z)$ and 2D view direction $(\theta, \phi)$ as input and outputs the outgoing radiance $\mathbf{c}=(r, g, b)$ and opacity $\alpha$ of a ray.
Similar to i-NGP, we adopt a hash encoder $\mathbf{E}_s$ to encode the input position to feature vector $\mathbf{E}_s(\mathbf{p})$ and use a shallow neural network $F_s$ to map the normalized view direction $\mathbf{v}$ in 3D Cartesian space to view-dependent color and opacity.
\begin{equation}
    F_s: (\mathbf{E}_s(\mathbf{p}), \mathbf{v})\rightarrow(\mathbf{c}_s, \mathbf{\alpha})
\end{equation}

Note that rendering such a continuous opacity field is quite efficient: we just need to query the network once for a ray.

\begin{figure}[t]
  \includegraphics[width=\linewidth]{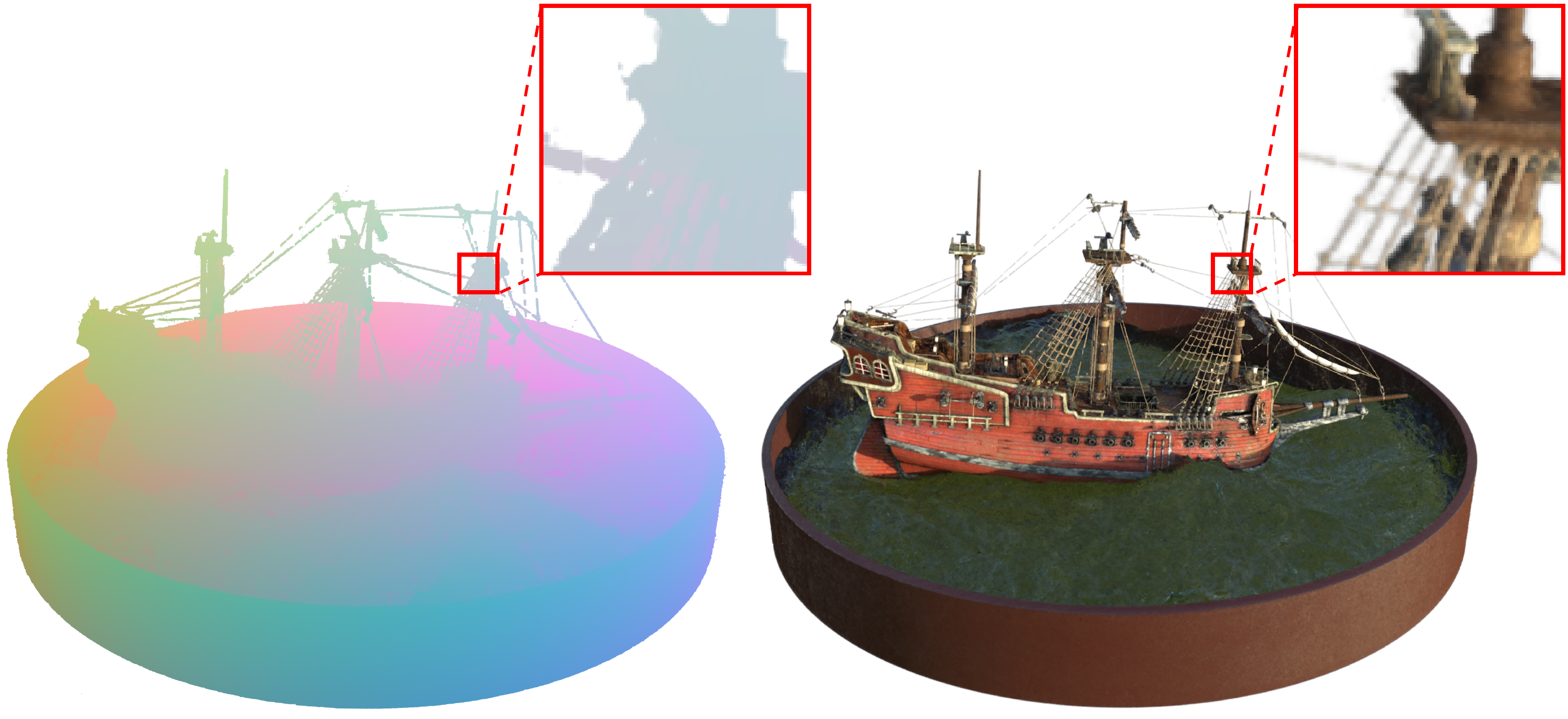}
  \caption{\textbf{The estimated hit point map (left) and rendered color image (right).} Though with the loss of accurate geometry details, our hit point estimation strategy can largely eliminate the ambiguity caused by proxy shape. Our i-NOLF can further recover both high-quality geometry and appearance details.}
  \label{fig:hitpoint_to_shading}
\end{figure}

\begin{figure*}[t]
  \includegraphics[width=0.95\linewidth]{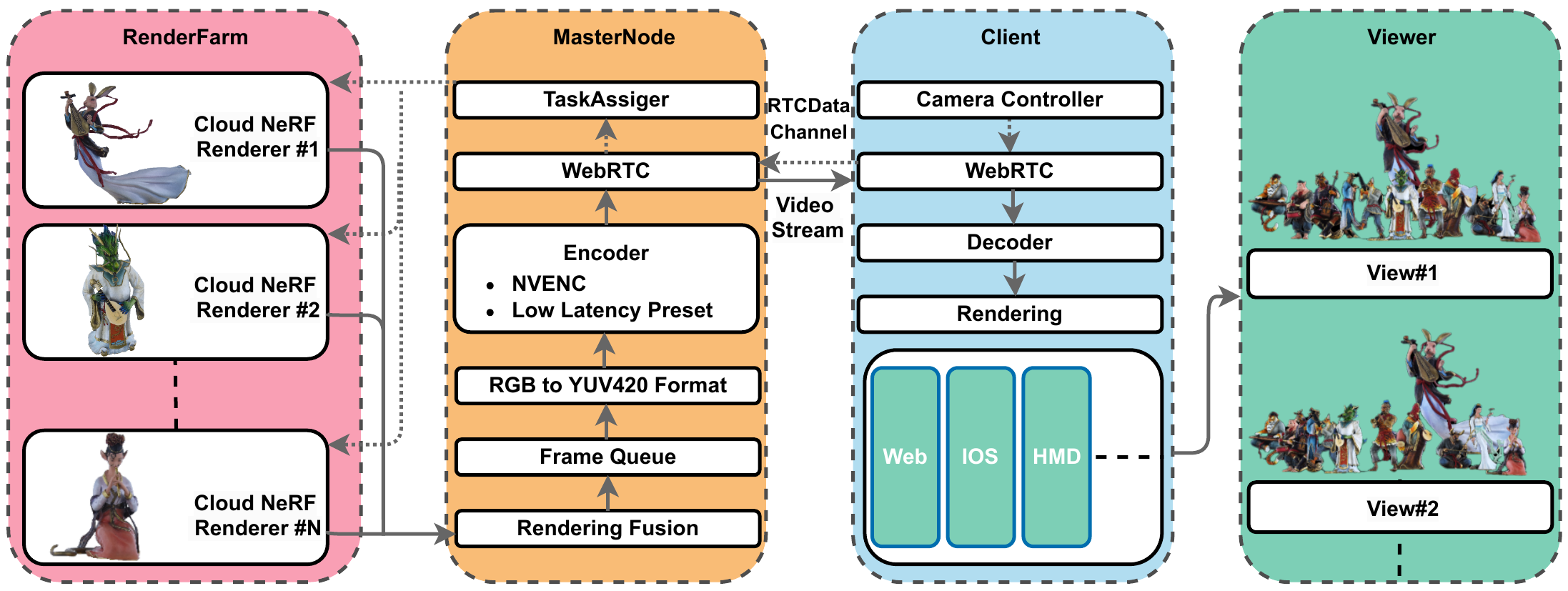}
  \caption{\textbf{Cloud Platform and data streaming illustration}.  To ensure the low latency and interactivity required by NEPHELE, we employ WebRTC as a streaming protocol to exchange information efficiently. When rendering a scene, the client-side camera controller sends user interaction data to the MasterNode of NEPHELE via RTCDataChannel every 20ms. Then the TaskAssigner in MasterNode assigns rendering tasks to the computational sub-nodes of the RenderFarm. The rendering results of all sub-nodes are synchronized to the rendering fusion module of the MasterNode for inter-scene fusion based on depth information. The rendered frames are compressed into a bitstream by using an NVENC encoder with low latency preset. Finally, the compressed bitstream is transmitted to the client through WebRTC for decoding and rendering.}
  \label{fig:Cloud Platform}
\end{figure*}

\paragraph{Hit Point Estimation and Opacity Refinement.} 
It is essential to keep the lumispheres of spatially adjacent sample points as similar as possible for better approximating the opacity light field using the instant neural network. Imagine the worst case to learn an opacity light field with the bounding box of an object as proxy geometry, The spatially encoded feature vectors of two adjacent points are similar, but the lumispheres are quite different. It is hard to approximate such almost one-to-many mappings using a shallow MLP.

To this end, we seek to exploit the geometry priors from i-NGP's density fields to estimate the ray-object hit point.
Recall that the original NeRF~\cite{mildenhall2021nerf} estimates the hit depth of a ray by blending the samples' depth with weights:
\begin{equation}
\begin{aligned}
    w_i &= T_i (1 - exp(-\sigma_i \delta_i)), \\    
    T_i &= exp(-\sum_{j=1}^{i-1}{\sigma_j \delta_j})
\end{aligned}
\end{equation}
where $\sigma_i$ is the density and $\delta_i$ is the distance between adjacent samples. Such an estimation scheme works well for solid surfaces but generates noisy hit points, especially on object boundaries and hair objects. To address this, we adopt an empirical strategy regarding the sample point with the largest blending weight $w_i$ to be the hit point in the ray marching process. Fig.~\ref{fig:hitpoint_to_shading} illustrates an estimated hit point map that helps optimize the neural opacity radiance fields.

In the meanwhile, for ease of opacity approximation, we accumulate the sample weights into the ray's coarse opacity $\alpha_c$, which is fed to $F_s$ and refined to the final opacity $\alpha$, analogous to ConvNeRF~\cite{convnerf}:
\begin{equation}
    F_s: (\mathbf{E}_s(\mathbf{p}), \mathbf{v}, \alpha_c) \rightarrow(\mathbf{c}_s, \mathbf{\alpha})
\end{equation}

\paragraph{Diffuse and Specular Component.} 
We further decouple the outgoing radiance function into view-independent diffuse component and view-dependent specular component, analogous to the idea of median removal~\cite{SLF} strategy, which effectively helps keep specular effects during surface light field compression. The key observation is that the specular component, which is the residual radiance after subtracting the rich diffuse component, should be more compressible~\cite{SLF} and learnable by neural network~\cite{DSLF}.

Similar to RefNeRF~\cite{verbin2022refnerf}, we adopt an extra spatial hash encoder $\mathbf{E}_d$ and neural network $F_d$ to encode the diffuse radiance function which takes only position as input and predicts diffuse color $\mathbf{c}_d$ and a specular tint $s$, 
\begin{equation}
    F_d: \mathbf{E}_d(\mathbf{p})\rightarrow (\mathbf{c}_d, t)
\end{equation}
The outgoing radiance can be obtained by combining them in the form of
\begin{equation}
    \mathbf{c} = \mathbf{c}_d + t \cdot \mathbf{c}_s
\end{equation}

\paragraph{Training.} 
To jointly learn the parameters of hash grids $\mathbf{E}_s, \mathbf{E}_d$ and neural networks $\mathbf{F}_s, \mathbf{F}_d$, we exploit $\mathcal{L}^2$ loss to constrain both radiance and opacity to be similar to the ground truth color $\mathbf{c}_{gt}$ and alpha values $\mathbf{\alpha}_{gt}$. 
\begin{equation}
    \mathcal{L} = \frac{1}{N}\sum_i^N(\|\mathbf{c} - \mathbf{c}_{gt}\|^2 + \|\alpha - \alpha_{gt}\|^2)
\end{equation}
where $N$ is the number of rays in a batch that are sampled according to an optimizable error map. We apply Adam optimizer to optimize our neural opacity light field with the same parameter settings as i-NGP.

\paragraph{Density Texture Cubes.}
Note that the hit point estimation process is required for both effective training and rendering, which means multiple time-consuming network queries are necessary to calculate the densities of sample points during the ray marching process. It leads to slower training speed and slower rendering speed compared to i-NGP with additional neural opacity light field query overhead. 
To address this issue, a straightforward way is to cache the continuous density field as a dense density grid for fast fetching, but the high memory footprint limits the grid resolution. In contrast, we represent the density field as a set of small dense density texture cubes to skip the empty space for compact storage. 
Specifically, we first build a base uniform indexing grid of low resolution, e.g., $b = 128$, to index a query that falls in a valid grid cell to the corresponding address in a table of valid cubes. Each cube in the table is actually, again, a uniform grid of resolution, e.g., $r = 16$, which stores the queried densities.
With the cached density cubes, the hit point estimation process (mainly the ray marching) is memory intensive and thus, we further boost it by moving the cubes to texture memory for higher cache coherency and exploiting the hardware-accelerated linear interpolation. 
So far, with the density texture cubes, we can obtain accurate ray-object hit points in less than $0.1ms$ for rendering an image of $1920\times1080$ resolution.

\paragraph{Diffuse Texture Cubes.}
Due to the decoupled network design, utilizing two independent networks separately to encode diffuse and specular components, once trained, we can further cache the diffuse network's output to diffuse texture cubes similar to the density texture cubes. Note that different from the density texture, we only cache the diffuse texture of the voxels surrounding all the possible hit points to reduce memory usage. In this way, we can achieve another rendering speed boost by skipping querying the network twice to avoid an additional time overhead. 

\paragraph{Implementation with PSH encoding.} 
Based on the fact shown in Table.~\ref{tab:PSH} that the PSH encoding can achieve superior memory efficiency and rendering speed at the cost of slightly degraded rendering quality, here we describe our implementation scheme to balance the pros and cons. 
Similar to the strategy to cache diffuse texture, we build the initial hash table to only store the features of voxels surrounding the hit points. Thus we can exploit finer grid resolution for better modeling capability with an acceptable parameter scale.  
Besides, since the diffuse component can be cached and only the specular network will be queried online, we only adopt PSH encoding for specular encoder $\mathbf{E}_s$ while keeping the hash grid for the diffuse encoder. This is also consistent with the motivation for decoupling the diffuse and specular components that the rich diffuse texture requires higher modeling capability.

\section{Cloud Platform} \label{sec:platform}
\subsection{Master renderer and composition}
Traditional NeRFs are running on the user's host device, which requires users to provide high-end GPU devices for NeRF rendering. 
And this solution only supports a single scene rendering at each time, which greatly limits the NeRF rendering experience.
Our goal is to make NeRF's rendering reach every user without requiring high-end GPU devices.
To this end, we propose a cloud rendering system consisting of a master node that deploys NeRF task scheduling and a sub-node that provides high-performance computing forming the NeRF cloud platform.
In addition, all NeRF subjects on the cloud platform can be freely composited to build a NeRF scenario.
The computation sub-nodes provide the color and depth information rendering of each individual NeRF subject, and the rendering results of all sub-nodes are synchronized to the master node.
Then the master node performs fusion between scenes according to the depth information.
Finally, the rendering results are streaming to the user's multi-display terminal through our streaming strategies.

 \subsection{Streaming strategies}

For WebRTC peer-to-peer communication, we implement a signaling server to exchange media and network metadata to bootstrap a peer connection between the cloud and the client.
When a client sends a request to the signaling server to experience the cloud rendering service, the signaling server assigns a MasterNode with service resources in NEPHELE for the client. 
A connection is established through a discovery and negotiation process. 
Then the client-side camera controller module frequently samples the user's position and orientation and converts the position and viewing direction into  an affine transformation matrix.
We transmit this matrix directly from the client to the MasterNode by using WebRTC's RTCDataChannel \cite{datachannel} every 20ms.
Since there is no intermediary server and fewer hops to transfer data using RTCDataChannel, it allows for lower transmission latency. 
With the matrix, the MasterNode assigns rendering tasks to the target GPUs in RenderFarm and composes the ray-based rendering frames in the correct order for the client.
The rendered frames are finally sent to the encoder to generate encoded packets.

NVIDIA GPUs contain a hardware-based encoder (referred to as NVENC) that provides fully accelerated video encoding \cite{NVENC}, so we use the H.264 codec implemented by NVENC as the encoder equipped with the MasterNode.
With complete encoding offloaded to NVENC, the graphics engine and the CPU are free for other applications, such as rendering.
For a typical end-to-end streaming scenario to incur low latency, we also minimize the encoding and decoding latency by choosing GOP with IPPP structure, no look-ahead, and the lowest possible VBV buffer for the given bitrate and available channel bandwidth. 
In the IPPP coding structure (no B frames), all of the P frames are forward predicted without backward or bidirectional prediction, which will make lower encoding and decoding latency, so it is suitable for our case.
As the rendered frames are in RGB format, we transformed them into the YUV420 format for efficient compression.
The compressed bitstream is packaged into RTP packets, encrypted, and transmitted to the client via WebRTC with low latency, which enables the client to receive the desired frames in real time after the interaction.
 The received frames can be decoded and displayed immediately without the need for a buffer.

\section{Scheduling for Multi-user} 
\label{sec:scheduling}
We discuss scheduling policies that MasterNode applies to distribute multiple rendering jobs in a group of Heavy and Light tasks for RenderFarm. The policy is for two major rendering scenarios: a single user viewing multiple objects in a re-defined scene and multiple users viewing the same object from adjacent angles.  
\subsection{Single-user-multi-object Scheduling}

The simplest solution to provide viewing services for a single user with a large scene containing numerous objects is to render each object independently and composite them into the final view. In this case, MasterNote transfers the user's camera view to all the NGP models (thereafter renderers) in RenderFarm, makes each renderer generate frame images and depth on its own, gathers them, and composites them into a large scene, then sends the results back to the user. This solution, however, encounters two obstacles: 1) the  I/O bandwidth limits of MasterNode can not handle simultaneous image transferring. It leads to unacceptable rendering response time for a single user viewing multiple objects in a large scene. 2) the workload on different GPUs in the RendFarm is unbalanced because the number of rays (denote as \textit{nhit}) hit by each NGP-NOLF object in a large scene is varied, requires a diverse volume of rendering resources. Simply assigning uniformed rendering resources to different objects causes wastes of GPU power for smaller-sized images and long tail latency for larger-sized ones.

We introduce a load-balancing schedule policy to separate objects into two groups-- smaller-sized and larger-sized. MasterNode assigns Larger-sized objects to HeavyFarm so that each object occupies one graphics card for better performance and distributes a group of smaller-sized objects LightFarm with shared GPU resources for higher utilization. The overall time spent on rendering a frame with multiple objects $T_{frame}$ can be expressed as follows:
\begin{equation}
\begin{aligned}
  T_{frame} = \max_{i \subset O}{(T_{i})} + \sum_{i\subset O} \frac{{D_{i}}}{BW} + C
\end{aligned}
\end{equation}
where $T_i$ is the rendering time, $D_i$ is the result data size of the NGP-NOLF instance $i$ that is related to the image resolution, $BW$ is the bandwidth limits of MasterNode, $C$ is the time necessary to consolidate all rendered results, and $O$ is the collection of rendered entities. 
\begin{figure}[t]
  \includegraphics[width=\linewidth]{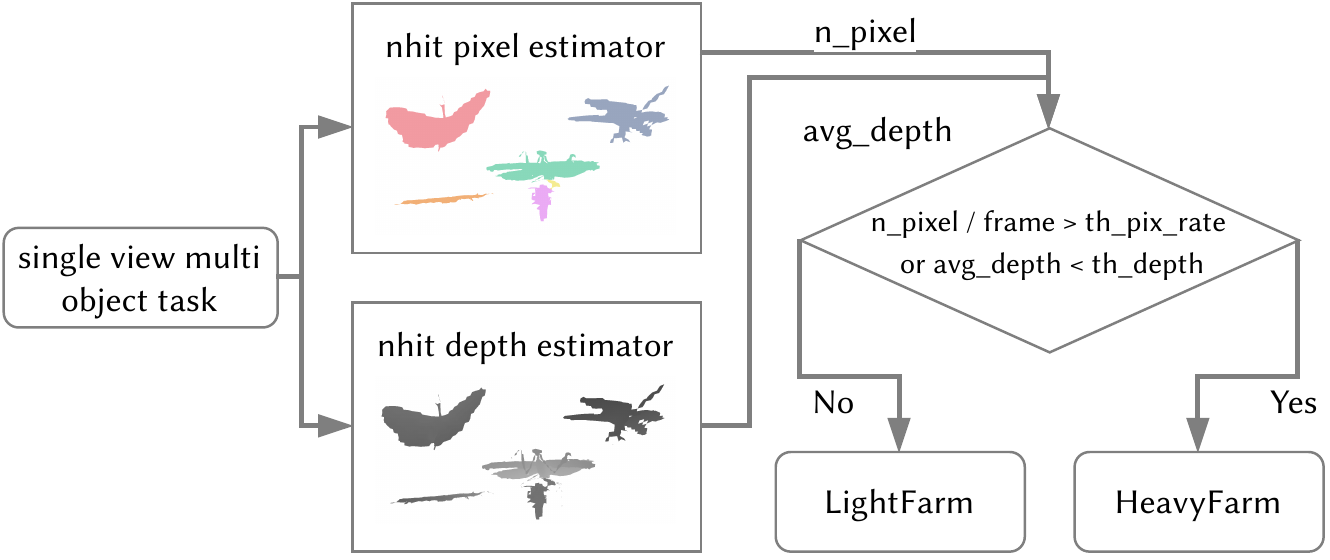}
  \caption{\textbf{Multi-object Scheduling} A flowchart depicts the process of scheduling multiple objects. Upon a single view multiple object task, two estimators in the scheduler calculate the pixel value (\textit{n\_pixel}) and the average depth (\textit{avg\_depth}) of \textit{nhit}, then pass them to a decision module. Tasks that need higher pix\_rate or lower depth will be assigned one or more dedicated graphics cards in HeavyFarm. The rest of the tasks run in LightFarm.}
  \label{fig:flowchart}
\end{figure}
The goal of the scheduler is to ensure a lower $T_{frame}$ while retaining the pre-defined frame resolution (represented as $D_i$). The scheduler should decrease the number of instances in $O$ as much as feasible and choose a proper subset in RenderFarm. We use a preconceived mesh to estimate the pixel quantity and depth on the final rendered screen to predict the number of corresponding \textit{nhit}. With the metrics, we can determine the necessity of rendering tasks and the target RenderFarm. Fig. \ref{fig:flowchart} demonstrates the fundamental idea of the scheduler.

We use EGL to generate a rough prior scene and render it off-screen as a \textit{nhit} estimator, which operates so quickly that its impact on rendering time is negligible. We employ Infiniband and CUDA-aware MPI for cross-node communication, and we can exploit the GPU Direct RDMA optimization on the Nvidia Advanced GPU to expedite the readout of the rendering results, therefore lowering the cross-node overhead.

\subsection{Multi-user Scheduling}
As for the multi-user scenario, the naive approach is to render each user's job in a round-robin manner to balance fairness and average response time. But it introduces repetitious rendering labor if users look at the same object from adjacent angles because the angles share a decent number of rays for rendering. Besides, the fairness policy of the round-robin has its limits to meeting the rendering response time for every user, which leads to unaccepted frames per second (FPS) for some users. We propose a multi-user scheduling scheme to improve render efficiency and meet the FPS challenge for the users.
\begin{algorithm}[t]
  \caption{NEPHELE Scheduling Policy}
  \label{pseudocode:multiuser}
  \providecommand{\RenderFarm}{\textsf{RenderFarm}}

  \SetKwData{TargetTime}{target\_time}
  \SetKwData{CurrentTime}{current\_time}
  \SetKwData{Loaded}{loaded\_tasks}
  
  \SetKwArray{Users}{users}
  \SetKwArray{Starved}{starved\_users}
  \SetKwArray{Rays}{rays}
  
  \SetKwFunction{Sort}{sort}
  \SetKwFunction{Filter}{filter}
  
  \While{\RenderFarm{} is available}{
  \For{$u \in$ \Users}{
    $u.\Loaded \gets$ \emph{\# of tasks loaded}\;
    $u.\TargetTime \gets T_\emph{last frame} + T_\emph{wait}$\;
  }
      \Sort{\Users, by \Loaded and \TargetTime}\;%
  
      \BlankLine
      \tcp{resolve starvation first}
      \Starved $\gets$ \emph{\Users whose $\TargetTime \leq \CurrentTime$}\;
  
      \While{\Starved $\neq \emptyset$}{
          deny \emph{requests from new clients}\;
          \For{$u \in \Starved$}{
              \emph{render all tasks of $u$ on \RenderFarm{}}\;
              \emph{update render time for $u$}\;
          }
          allow \emph{requests from new clients}\;
    }
  \BlankLine
  \tcp{render until \RenderFarm{} full}
      \Rays $\gets$ \emph{max supported rays on \RenderFarm{}}\;
      \For{$u \in \Users$}{
          \lIf{\Rays not enough to render $u$}{
              break
          }
          \For{$t \in$ shared tasks of $u$}{
              \emph{render $t$ on \RenderFarm{}}\;
              \emph{update render time for all users of $t$}\;
              $\Rays \gets \Rays - \emph{rays of $t$}$\;
          }
          \For{$t \in$ dedicated tasks of $u$}{
              \emph{render $t$ on \RenderFarm{}}\;
              \emph{update render time for $u$}\;
              $\Rays \gets \Rays - \emph{rays of $t$}$\;
          }
      }
  }
\end{algorithm}
\subsubsection{Improve render efficiency}
We first discuss how to optimize the response time ($T$) for rendering multiple jobs from users that are looking at the same object. We have
\begin{equation}
\label{eq:rendertime}
\begin{aligned}
  T ={} & \left(\left(k T_{M_{S}} + T_{N_{S}}\right) \times n + \sum_{i = 0}^{n} \left(k T_{M_{i}} +  T_{N_{i}}\right) \right)
\end{aligned}
\end{equation}
where $k$ is a constant coefficient, $T_M$ is the time for mesh rendering, $T_N$ is the time for i-NOLF rendering, and $n$ is the number of users requesting rendering services. Note that the $S$ refers to the shareable rays for adjacent angles and global view, and $i$ is the necessary rays to render a specific view angle. Hence the union of $S \times n$ and $\sum{i}$ is the total number of rendering jobs submitted. Since the efficiency $E$ is the inverse of $T$, we can observe that the rendering efficiency can take significant advantages from the more shareable rays by maximizing the number of shareable rays. 

TaskAssigner in MasterNode treats the shareable rays as Heavy Tasks and places all of them HeavyFarm for better performance. The necessary rays for specific viewing angles go to whichever graphic card is available in LightFarm. In this case, multiple necessary rays for different angles can render in parallel. In this way, Eq. \ref{eq:rendertime} can be expressed as:
\begin{equation}
\label{eq:rendertime_max}
\begin{aligned}
  T ={} & \left(\left(k T_{M_{S}} + T_{N_{S}}\right) + \max_{i = 0}^{n} \left(k T_{M_{i}} +  T_{N_{i}}\right) \right)
\end{aligned}
\end{equation}

TaskAssigner also maintains a historical table to cache the re-usable tasks and guide RenderFarm to render the shareable rays offline in advance. In this case, Eq. \ref{eq:rendertime_max} can be further optimized as:
\begin{equation}
\label{eq:rendertime_offline}
\begin{aligned}
  T ={} & \max_{i \subset n} \left(k T_{M_{i}} +  T_{N_{i}}\right)
\end{aligned}
\end{equation}

\subsubsection{Meet the FPS challenge for users}
Besides the rendering efficiency, our system offers the best-effort service for achieving the requested FPS by preventing rendering an excess number of users, as in Algorithm \ref{pseudocode:multiuser}.
TaskAssigner assigns a wait time $w_i$ for each user $i$, and we have:
\begin{equation}
w_i = \frac{1}{\mathrm{FPS}_i} - t_i
\end{equation}
Where $\mathrm{FPS}_i$ is the target FPS, and $t_i$ is the moving average of job completion time in seconds. $w_i$ is updated upon the completion of a render task. A task 
is starved if has not started in $w_i$ after the last frame. When starved task exists, TaskAssigner will be in a starvation state and deny requests from
new clients until all starved tasks are resolved.

\begin{figure*}[t]
  \includegraphics[width=1.0\linewidth]{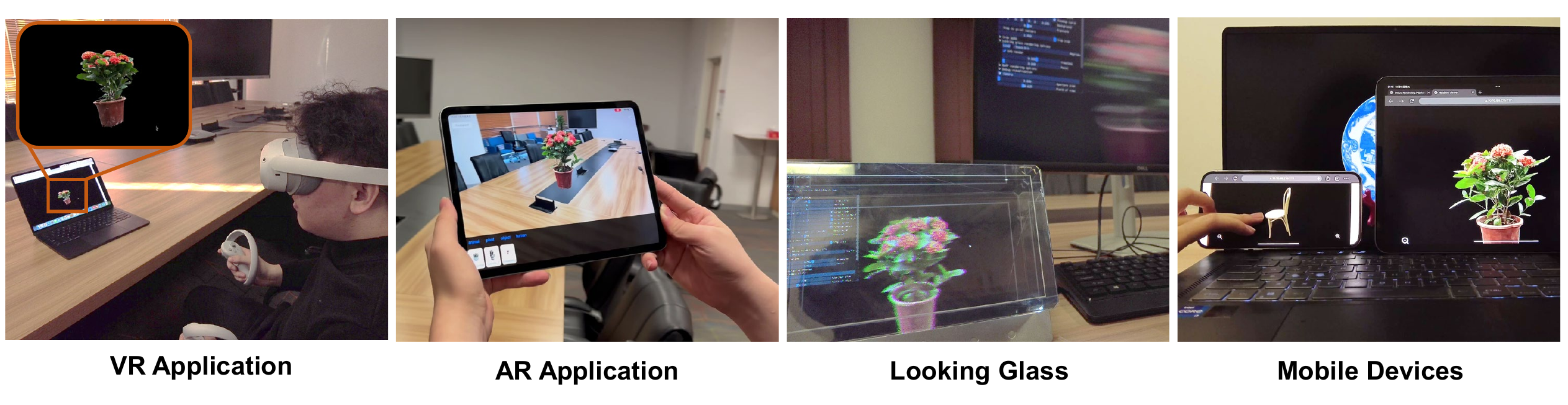}
  \caption{\textbf{Illustration of NEPELE application, including VR application, AR application, rendering on Looking Glass holograms display, and Mobile devices}. }
  \label{fig:applications}
\end{figure*}

\section{Applications and Demonstrations} \label{sec:application}
\begin{figure*}[t]
  \includegraphics[width=1.0\linewidth]{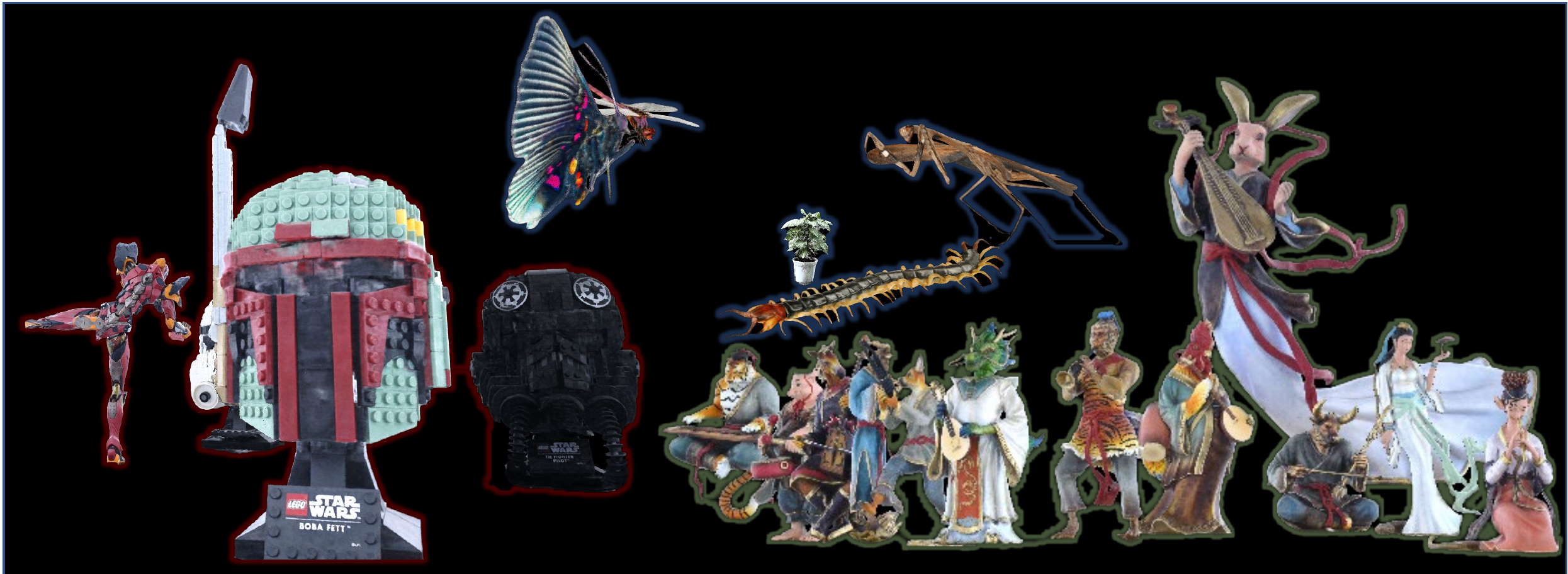}
  \caption{\textbf{Illustration of NEPELE application of multi-user and multi-object scenarios, such as Lego Star War (red highlight), Insects Fighting (blue highlight), and Chinese Zodiac (green highlight)}.}
  \label{fig:multiple}
\end{figure*}

Benefiting from such a neural cloud rendering platform, we further demonstrate our cloud radiance rendering capabilities through a range of applications, from cloud VR/AR rendering, to sharing NeRF assets between multiple users on mobile devices (as shown in Fig.~\ref{fig:applications}), and allowing NeRF assets to be freely combined into a new scene (as shown in Fig.~\ref{fig:multiple}).
\paragraph{VR Application.}

Head-mount displays (HMD) provide an immersive VR experience for users. Our solution, including the frontend and backend, enables users to interactively enjoy VR streaming directly in the browser. Our frontend adopts the web framework A-Frame to support any WebXR-supported browsers. By registering as an A-Frame system, our frontend establishes a WebRTC connection with the backend after we detect an HMD. After the connection is established, we continuously acquire the user's view and pose from the WebXR camera with a minimum interval of 20~ms. The view and pose are converted by Three.js into the same matrix used in Instant-NOLF, then send to the backend by RTCDataChannel. The backend renders the content according to client data into a stereo frame, which is transmitted to the client via WebRTC. The frontend will decode the video, and split the video into two views by changing uv attribute in the mesh element of the A-Frame video primitive.

\paragraph{AR Application.}

Our streaming-based solution enables users to interactively experience high-fidelity and photorealistic AR on their iPad or mobile phone over a wireless network, allowing them to project a pot of flowers in front of them. Users can break the boundaries between the real and virtual worlds and appreciate the flowers from any angle in an immersive way. The implementation details of our approach under the AR setting are as follows:
The mobile device adopts the ARKit's ARCamera to continuously acquire the camera's intrinsic matrix and transformation matrix that defines the camera's rotation and translation in world coordinates. When the camera detects the plane, we also get an ARPlaneAnchor, which records the center and extent of the plane in the anchor's coordinate space. We transmit these matrices and ARPlaneAnchor directly from the mobile device to the cloud server every 20ms by using WebRTC's RTCDataChannel. The cloud server renders the content with a pure green background according to the parameters from mobile and encodes the rendered frame into a bitstream by using an NVENC  encoder with low latency preset. Then the bitstream is transmitted to the mobile device via WebRTC with low latency. After receiving the bitstream, the mobile device decodes the video frames and uses the chroma keying matting method to make the background of the decoded frames transparent. Finally, we combine the real and virtual worlds to create an immersive experience by overlaying the frames onto the camera screen of ARSCNView.

\paragraph{Looking Glass Display.}
Looking glass is a 3D holographic displayer that needs 45 horizontally consecutive camera poses to render a looking glass scene. 
The goal of the camera is to emulate what the human eye would see if the 3D scene were real in physical space. 
The easiest way to think of this is to imagine that the screen at the base of the Looking Glass was a window pane: a flat rectangular portal through which we're viewing the 3D objects. 
Rendering 45 views at the same time is a challenge for existing methods.
We tried it on i-NGP, and with our method, the rendering speed of i-NGP becomes 2-5 FPS.
Obviously, there will be a lot of redundant rays that can be reused between the 45 views. 
Our approach is very suitable for this rendering mode, which achieves a much faster speed of up to 10-20+ FPS. 
\begin{figure*}[t]
  \includegraphics[width=1.0\linewidth]{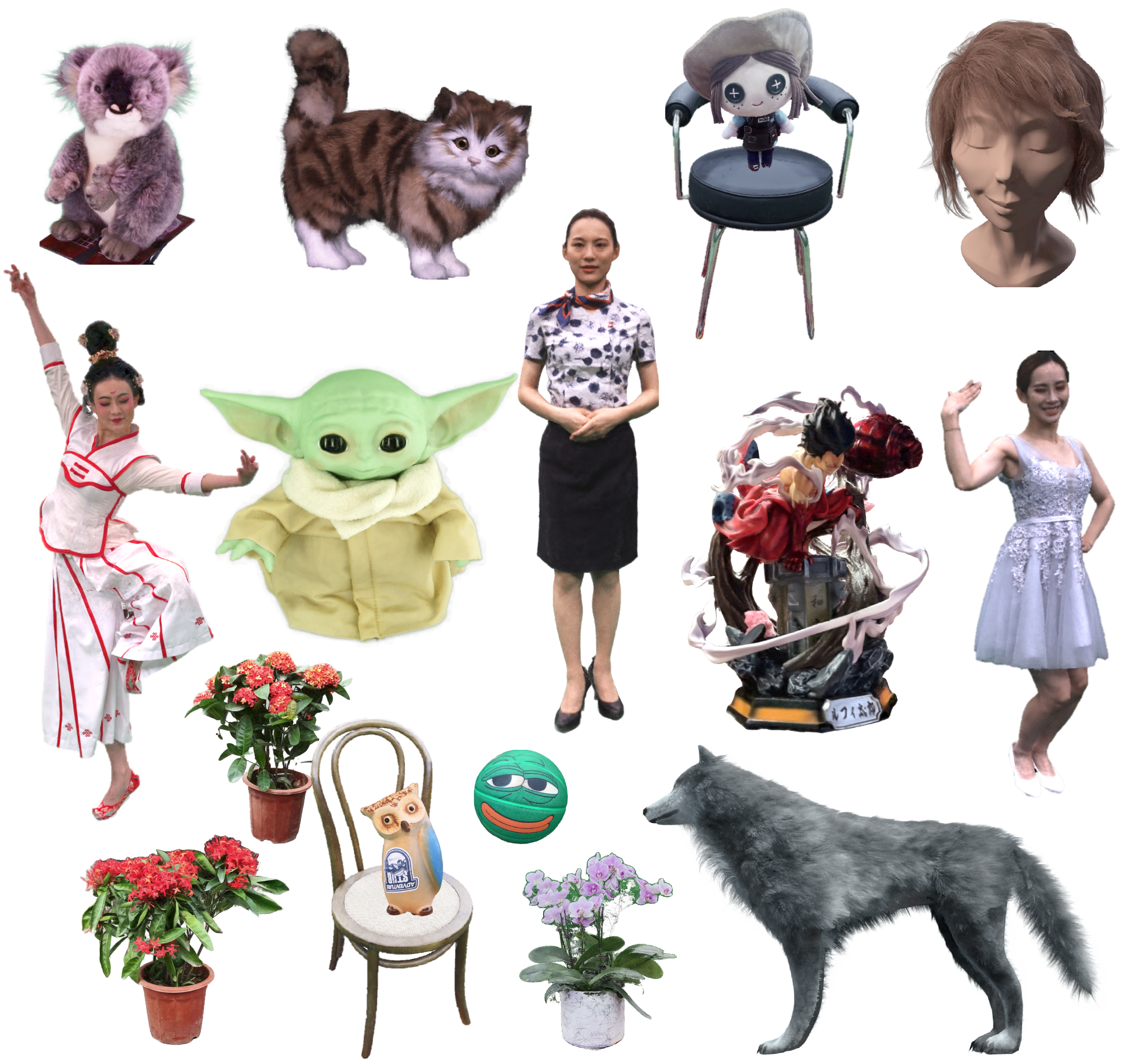}
  \caption{\textbf{Gallery of our synthetic results}. Our method can generalize well to various objects, including objects with a solid surface, complex geometry, and fuzzy objects, such as humans, toys, animals, plants, furniture, and sculptures.}
  \label{fig:Gallery}
\end{figure*}

\paragraph{Cloud Rendering On Mobile Devices.}
Our proposed NEPELE platform also supports various mobile devices, which frees NeRF techniques from the need for high-performance computing devices and makes NeRFs available for everyone. 
We run our NEPELE platform on our computing clusters, and users can experience NeRF scenes using web viewers on their Phones, Pads, and Laptops by connecting to our cloud rendering platform service.
Cloud rendering speed on mobile devices for a single user can reach 25+ FPS depending on the user's network bandwidth.
All our results can be experienced on our cloud rendering platform using mobile devices without the need to download additional model data.

\paragraph{Cloud Rendering For Multi-user Scenarios}
The problem with traditional NeRF based methods is that they run independently on a local computing device. 
Therefore, they do not have the feature of sharing a NeRF scene representation among multiple people and the combination of NeRF scenes.
Our proposed NEPELE platform provides a large number of NeRF scenes, which not only supports multiple users to experience a certain NeRF scene but also allows NeRF scenes to be freely combined into a new scene.
NEPELE benefits from the high-performance computing cluster provided by the remote cloud as well as the uniquely designed Instant-NOLF and ray-level scheduling algorithm, which enables the real-time rendering of multiple NeRF combined scenes even under multiple users.

\section{EXPERIMENTS}\label{sec:results}

In this section, we evaluate our i-NOLF in various challenging scenarios. We first report the datasets adopted to evaluate our approach and the current state-of-the-art (SOTA) neural rendering approaches against which our method is compared. Then we provide the comparisons of our model with the baseline models for both rendering quality and rendering speed. It demonstrates while maintaining comparable excellent rendering quality, our method can achieve much higher rendering speed than the baselines, especially in high rendering workload scenarios (e.g., 4k resolution rendering), compared to i-NGP. We also perform an ablation study to verify the effectiveness of each component of our method. We further conduct a thorough speed-quality tradeoff analysis that provides key information for deploying our models to the cloud. Finally, the limitation and discussions regarding our system are provided in the last subsection.

\paragraph{Datasets.} 
We adopt the common benchmark dataset Synthetic-NeRF~\cite{mildenhall2021nerf} for fair comparisons with other methods. 
Besides, we capture various real objects to validate the generalization capability of our method in different scenarios. Specifically, we use a greenscreen dome with 80 cameras to capture several humans wearing various garments in complex poses, and we take 100-200 images of some common objects of various materials and shapes, such as potted plants, sculptures, furniture, and dolls, with a mobile phone. We exploit the off-the-shelf background matting~\cite{lin2021real} to obtain the foreground alpha mattes. We uniformly select 10\% images for testing and others for training. 
Finally, we further adopt the fuzzy-objects dataset from ConvNeRF~\cite{convnerf} to demonstrate the capability of our method on objects with extremely complex geometry, such as "wolf" and "hair".
We illustrate part of our representative rendering results in Fig.~\ref{fig:Gallery}.

\paragraph{Baselines.}
We adopt the original NeRF~\cite{mildenhall2021nerf} as the baseline model. Besides, we compare our method with several most recent fast neural rendering techniques, including DVGO~\cite{sun2022direct}, TensoRF~\cite{chen2022tensorf}, and approaches that exploit explicit acceleration data structure and custom CUDA codes such as PlenOctrees~\cite{yu2021plenoctrees}, PlenOxels~\cite{yu_and_fridovichkeil2021plenoxels} and i-NGP~\cite{NGP}. We also compare with MobileNeRF\cite{chen2022mobilenerf}, which similarly aims to achieve radiance rendering on mobile devices, but differently adopts explicit meshes and textures for local rendering via traditional OpenGL pipeline.

\begin{figure*}[t]
  \includegraphics[width=1.0\linewidth]{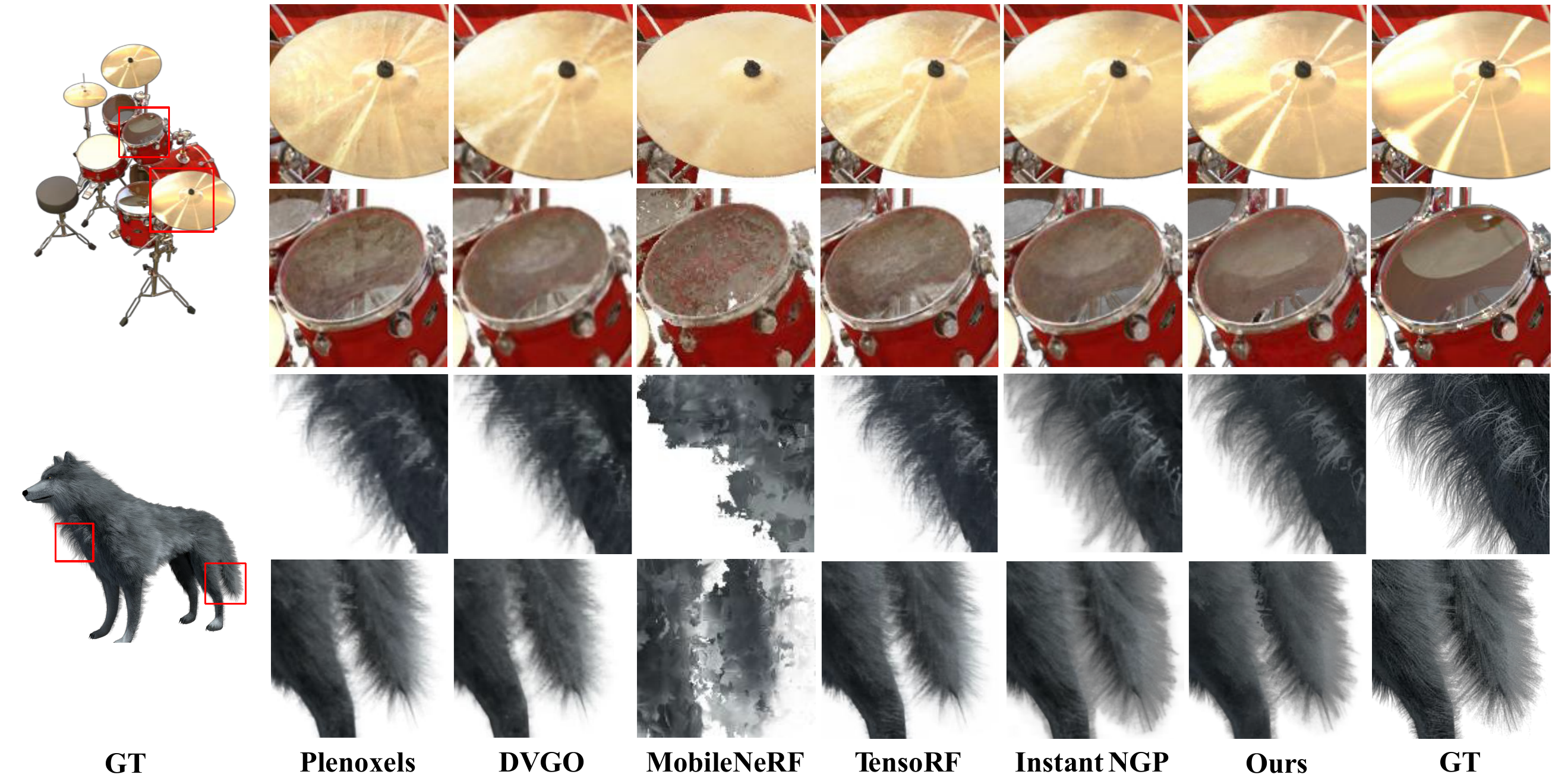}
  \caption{\textbf{Qualitative comparison of free-view synthesis}. Our method can better recover specular effects and fur details.}
  \label{fig:comparisons}
\end{figure*}

\setlength{\tabcolsep}{12pt}
\begin{table*}[t]
\centering
\caption{\textbf{Quantitative comparisons on each dataset}. We report average metrics of baselines and our approach. Overall our method achieves the best rendering quality.}
\resizebox{0.85\linewidth}{!}{
\begin{tabular}{lccccccccc}
\toprule%
\multirow{2}{*}{Method} & \multicolumn{3}{c}{Synthetic-NeRF} & \multicolumn{3}{c}{Real-Captured } & \multicolumn{3}{c}{Fuzzy-Objects} \\ 
\cmidrule(r){2-4} \cmidrule(r){5-7} \cmidrule(r){8-10}
         &  PSNR $\uparrow$    & SSIM $\uparrow$      &  LPIPS $\downarrow$      & PSNR $\uparrow$      & SSIM $\uparrow$      & LPIPS $\downarrow$      &  PSNR $\uparrow$     & SSIM $\uparrow$      & LPIPS $\downarrow$     \\ 
\midrule[0.3pt]
NeRF                 & 31.00          & 0.947        & 0.081        & 27.83      & 0.930      & 0.087      & 31.20      & 0.917      & 0.102     \\
PlenOctree           & 30.39          & 0.946        & 0.073        & 24.86      & 0.918      & 0.104      & 31.35      & 0.915      & 0.114     \\
PlenOxels            & 31.74          & \third{0.958} & \best{0.049} & 27.30      & 0.935      & 0.102      & 32.04      & \second{0.930}      & \third{0.100}     \\
MobileNeRF           & 30.89          & 0.947        & 0.062        & 27.02      & 0.902      & 0.108      & 30.82      & 0.911      & 0.110     \\
DVGO                 & 31.95          & 0.957        & 0.053        & \third{29.22}      & \third{0.938}      & \third{0.080}      & 33.69      & 0.925      & 0.106     \\
TensoRF              & \best{33.21}    &\best{0.963}  & \third{0.051}   & 28.60      & \best{0.948}      & 0.085      & \best{34.67}      & \best{0.948}      & \second{0.081}     \\
i-NGP          & \second{32.79}   & 0.920        & 0.061        & \second{29.48}      & 0.919      & \second{0.065}      & \third{34.35}      & 0.881      & 0.105     \\ 
\midrule[0.3pt]
Ours                 & \third{32.73} & \second{0.960} & \best{0.049} & \best{29.62} & \second{0.947} & \best{0.058} & \second{34.61} & \third{0.928} &  \best{0.073}   \\
\bottomrule%
\end{tabular}
}
\label{Tab: rendering comparisons}
\end{table*}

\begin{figure}[t!]
  \includegraphics[width=\linewidth]{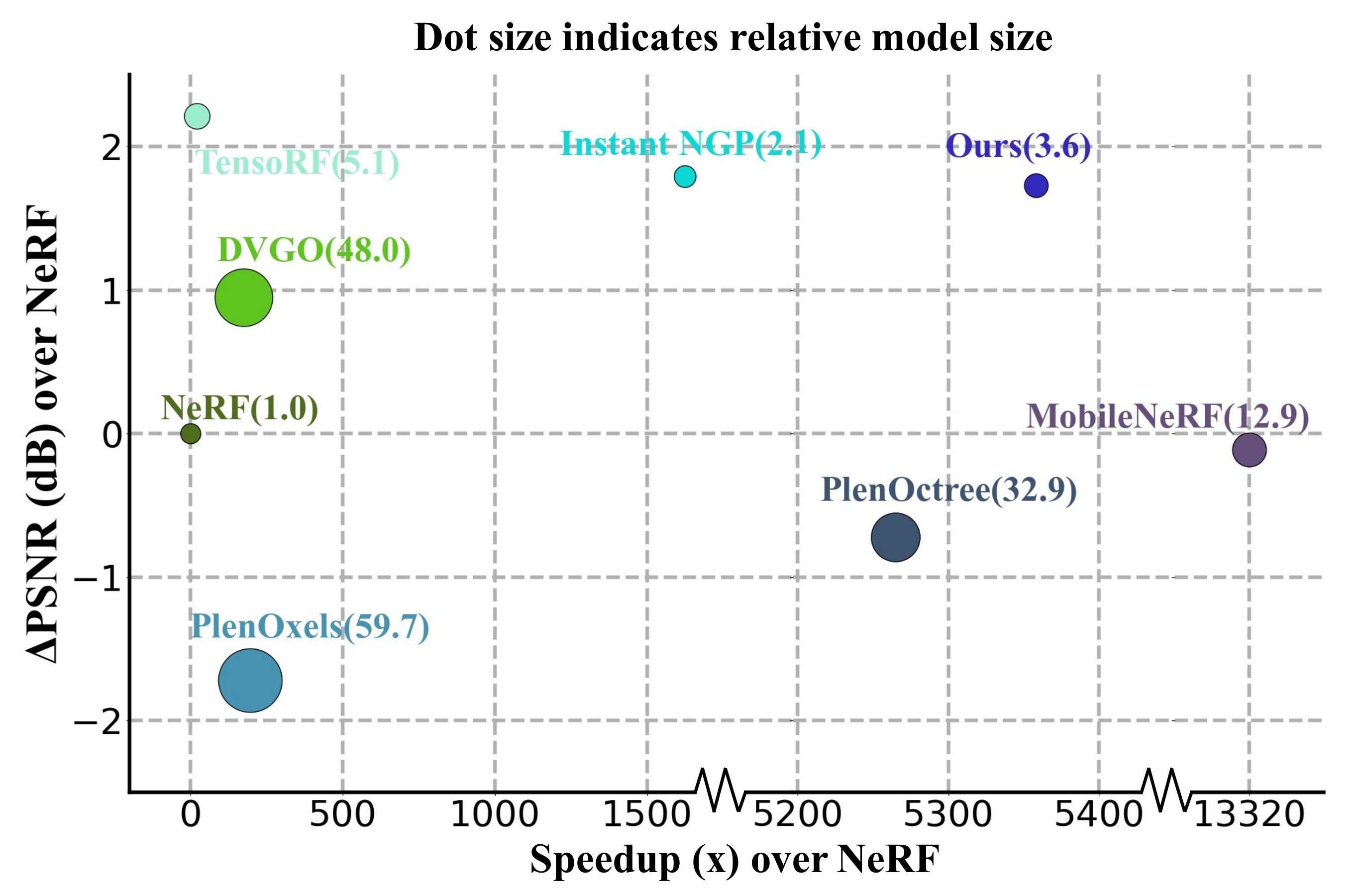}
  \caption{\textbf{Speedup-PSNR-Model Size comparisons.} Our method achieves a more favorable speedup-PSNR-model size tradeoff compared with other fast neural rendering methods. Note that dot size indicates relative model size, and our method achieves a combination of fast rendering speed, high-quality rendering, and small model size. }
  \label{fig:speed comparison}
\end{figure}

\setlength{\tabcolsep}{2pt}
\begin{table}[t]
\centering
\caption{\textbf{Quantitative comaprisons of rendering speed}. We report average metrics on the full NeRF synthetic dataset and model size and rendering speed at 800x800 resolution on the scene "Lego". Our method realizes an observably faster speed with high rendering quality.}
\resizebox{1\linewidth}{!}{
\begin{tabular}{l|cc|cccc}
Method & $\Delta$PSNR$\uparrow$ & $\Delta$SSIM$\uparrow$ & Size (MB)$\downarrow$ & Time (ms)$\downarrow$ & FPS$\uparrow$ & Speed-up$\uparrow$\\ \hline
NeRF            & 0.00 & 0.000 & 13.6 & 10417.28 & 0.09 & 1.00\\
PlenOctree      & -0.61 & -0.001 & 446.9 & 2.11 & 473.93 & 5265.89\\
PlenOxels       & 0.74 & 0.011 & 812.3 & 56.39 & 17.73 & 197.00\\
MobileNeRF      & -0.11 & 0.000 & 176.1 & 0.83 & 1198.86 & 13320.67\\ 
DVGO            & 0.95 & 0.010 & 652.8 & 63.49 & 15.75 & 175.00\\ 
TensoRF         & 2.21 & 0.016 & 68.8 & 488.23 & 2.04 & 22.67\\ 
i-NGP     & 1.79 & -0.027 & 28.7 & 6.84 & 146.18 & 1624.22\\ 
\hline
Ours            & 1.73 & 0.013  & 48.8 & 2.07 & 482.30 & 5358.89
\end{tabular}
}
\label{tab:speed comparisons}
\end{table}

\subsection{Rendering Quality Comparisons} \label{quality_compare}
Here we compare our approach with baselines for free-view rendering. For fairness, we optimize 200k iterations for baseline models as well as ours at the same original resolution. 
Fig.~\ref{fig:comparisons} provides qualitative comparisons against the baselines on the scene "drums" and "wolf". 
It's non-surprising that MobileNeRF fails to reconstruct the translucent surface of the drums since it adopts binary opacities to avoid sorting polygons. Besides, the limited capability of textured mesh leads to failure to model specular effects, and other methods, e.g., PlenOxels, tend to synthesize noisy images. On the contrary, our approach performs better in recovering the highlights favorably. 
For fuzzy objects, our i-NOLF can recover both clear geometry and appearance details, especially in the boundary region, which is essential for furry animals, e.g., By contrast, the baselines generate blurry fur details. Specifically, i-NGP suffers from obvious ghosting effects, and MobileNeRF fails to model such complex geometry limited by the meshes. 
For quantitative comparisons, we adopt Peak Signal-to-Noise Ratio (\textbf{PSNR}), Structural Similarity (\textbf{SSIM}), and Learned Perceptual Image Patch Similarity (\textbf{LPIPS})\cite{2018arXiv180103924Z} as metrics to evaluate the rendering quality of our model. We report the average metrics on each dataset for each approach in Table.~\ref{Tab: rendering comparisons}.  
Our approach outperforms most of the baselines, demonstrating that our approach can generalize well to various types of objects.

\subsection{Rendering Speed Comparisons} \label{speed_compare} 
Here we further evaluate the rendering speed of i-NOLF, which is essential to meet the interactive requirements in the cloud scenarios.
We report model size and rendering speed on the scene "Lego" and we also report the improved PSNR and SSIM compared to the original NeRF as a reference.
As shown in Tab.\ref{tab:speed comparisons}, our approach not only achieves an ultra-fast rendering speed but also achieves a better rendering quality. 
Specifically, TensoRF achieves the best rendering quality at slow rendering speeds, i.e., 2.04 FPS, while our method can render at 400+ FPS. PlenOctree realizes a similar rendering speed with a loss of rendering quality. 
Note that though MobileNeRF achieves a speed of 1000+ fps, it is a local renderer in need of multiple meshes and textures that requires a very long training time and a complicated preprocessing process, while the rendering quality is worse than NeRF. 
Besides, the large size of the meshes and textures leads to a long time before rendering to load data from the server for mobile devices, while our cloud solution can naturally achieve a seamless user experience.
As shown in Fig.\ref{fig:speed comparison}, our method achieves the best rendering quality-speed-model size tradeoff.

\begin{figure}[t!]
  \includegraphics[width=0.90\linewidth]{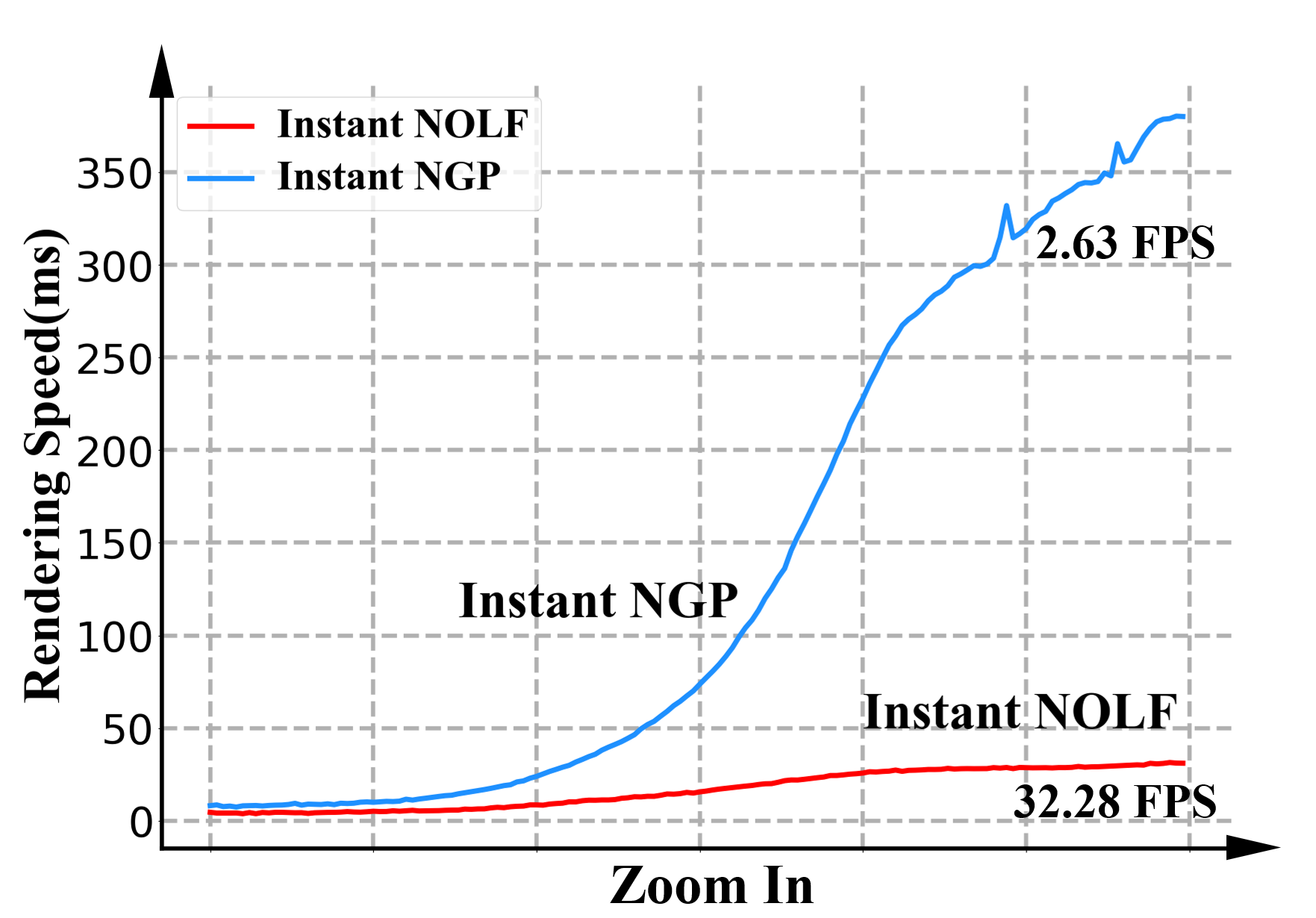}
  \caption{\textbf{Speed-Distance comparisons.} We make comparisons between our i-NOLF and i-NGP on rendering speed with the camera zooms in. While the rendering time of i-NGP increases in exponential order as the viewer gets closer, our i-NOLF still remains at an excellent speed, even at a close view position. }
  \label{fig:speed-distance comparison}
\end{figure}

Furthermore, we evaluate the speed of i-NGP and our i-NOLF for rendering heavy tasks by zooming in the camera. We render the scene "Lego" on a single Nvidia GeForce RTX4090 GPU at 4K resolution. As Fig.\ref{fig:speed-distance comparison} shows, i-NGP has similar rendering speeds with i-NOLF when the camera is far, but the rendering time of i-NGP increases exponentially to be unacceptable. 
On the contrary, the rendering time of i-NOLF increases almost linearly since it only requires one network query for rendering a pixel. 
Finally, when the rendered object fills the whole rendering screen, i-NGP comes to a poor FPS of only 2.63 FPS, while our i-NOLF still has about 32.28 FPS. It demonstrates the excellent rendering efficiency of our i-NOLF.

\subsection{Ablation Study} \label{ablation}
\begin{figure}[t!]
  \includegraphics[width=\linewidth]{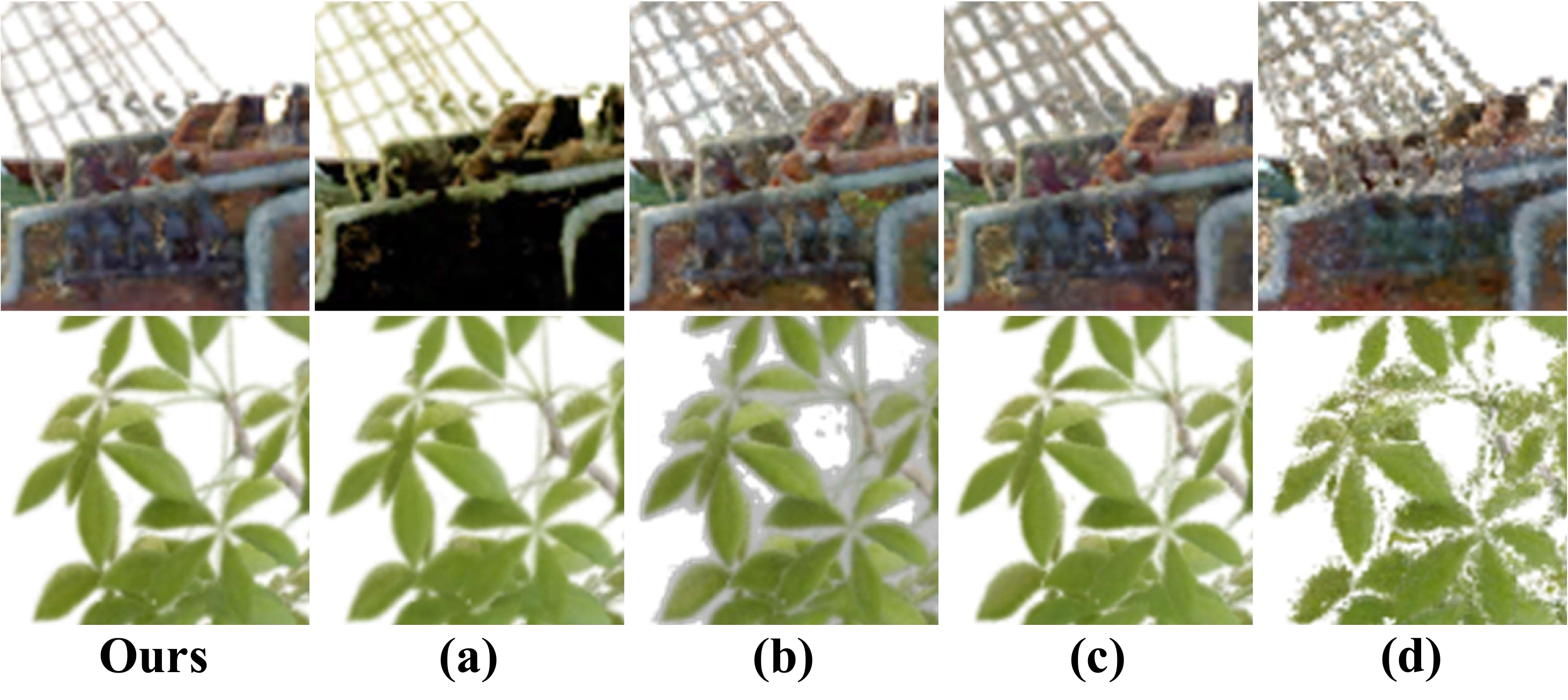}
  \caption{\textbf{Qualitative ablation studies.} We show the effectiveness of each component on the challenging parts of the "ship" scene and "ficus" scene: 
  (a) without diffuse component. (b) without view-dependent opacity. (c) without opacity refinement. (d) without hit point estimation.}
  \label{fig:rendering_ablation}
\end{figure}

We further conduct ablation studies of our approach on NeRF synthetic dataset, as illustrated in Table~\ref{tab:ablation}. We train a model ("w/o hit point") with a coarse proxy geometry slightly bigger than the real object, and the performance drops significantly, showing the effectiveness of the hit point estimation strategy helps approximate the neural opacity light field. 
Removing the view-dependent opacity output ("w/o opacity") severely degrades the rendering metrics since the cached density field cannot provide exactly the same geometry as the real object. Besides, directly predicting opacity without coarse opacity input ("w/o refine opacity") also reduces the rendering performance. 
Finally, replacing the specular tint with a constant value, e.g., 0.5, slightly decreases performance ("w/o tint"), and the model without diffuse component ("w/o diffuse color") shows the benefits of decoupling diffuse and specular components.
Fig.~\ref{fig:rendering_ablation} shows the visual details of different models. Without hit point estimation, the learned opacity light field is quite noisy due to the limited capability of shallow MLP. It fails to reconstruct objects with complex geometry such as "ficus" without predicting view-dependent opacity, while the opacity refinement strategy further helps recover the fine details.  

\subsection{Speed-quality Trade-off Analysis}\label{sq_analysis} 

While our i-NOLF achieves superior rendering speed, it's more important to analyze the speed-quality trade-off for deploying our model on the cloud. As shown in Table~\ref{tab:qs-trade-off}, we quantitatively analyze the impact of various components on speed, model size, and performance on a single Nvidia GeForce RTX4090 GPU. 
We report the average metrics for rendering quality on the full Synthetic-NeRF dataset. For rendering speed, we calculate the average runtime on the scene "lego" of rendering 200 random views in 4k resolution, simulating the high concurrent rendering workload in the cloud scenarios. In particular, we additionally report the runtime of the ray-marching process in which querying the network takes up most of the time.

In Table~\ref{tab:qs-trade-off}, firstly, the density texture (denoted as "dt") increases the rendering frame rate by five to six times compared to the base model, i.e., estimating the hit point by querying the density network. 
Using an MLP to approximate the diffuse component ("diffuse MLP") achieves better rendering performance at the cost of reducing the frame rate by 7 and doubling the model size. On the other hand, adopting the diffuse texture ("dift") almost eliminates the overhead for querying diffuse network, while slightly degrading the quality. 
Finally, we equip our model with perfect spatial hash encoding (PSH) instead of multiresolution hash encoding to further decrease around 20\% ray-marching time and 15\% total rendering time. Increasing the resolution of PSH encoding, e.g., from $128^3$ to $512^3$ slightly increases the rendering quality while the runtime is almost the same.
Note that since overall the ray-marching time of our model is reduced by at least 90\% by only querying the network once, the rest of the time to render an image, i.e., allocating memory for initialization, becomes a bottleneck for further optimization.
We also report the extra runtime memory required by density and diffuse textures without compression, noting that it is totally acceptable in cloud scenarios where multiple users share the same textures for a scene.

\setlength{\tabcolsep}{8pt}
\begin{table}[t]
\centering
\caption{\textbf{Ablation studies}. We evaluate each component of our model on the Syhthetic-NeRF dataset.}
\resizebox{0.8\linewidth}{!}{
\begin{tabular}{l|ccc}
Model                               & PSNR$\uparrow$ & SSIM$\uparrow$   & LPIPS$\downarrow$ \\ \hline
Ours, w/o hit point                 & 28.48          & 0.924            & 0.102 \\
Ours, w/o opacity                   & 23.38          & 0.919            & 0.088 \\
Ours, w/o refine opacity            & 32.48          & \best{0.960}     & \third{0.052} \\
Ours, w/o tint                      & \third{32.65}  & \best{0.960}     & \second{0.051} \\
Ours, w/o diffuse color             & \second{32.66} & \second{0.959}   & \third{0.052} \\ \hline
Ours, full                          & \best{32.73}   & \best{0.960}     & \best{0.049}
\end{tabular}
}
\label{tab:ablation}
\end{table}

\setlength{\tabcolsep}{3pt}
\begin{table}[t]
\centering
\caption{\textbf{Quality-speed trade-off analysis}. We evaluate the rendering quality of each model on the Synthetic-NeRF dataset. For rendering speed, we report the average ray-marching / total rendering time (in milliseconds) and frame rate to render the scene "Lego" at 4k resolution. We also report model size and extra required runtime memory (in megabytes) for caches which are important to deploy our models.}
\resizebox{1\linewidth}{!}{
\begin{tabular}{l|cc|cccc}
Model & PSNR$\uparrow$ & SSIM$\uparrow$ & Size$\downarrow$ & \makecell{Extra\\Memory}$\downarrow$ & Time$\downarrow$ & FPS$\uparrow$\\ \hline
i-NGP                 & 32.79 & 0.920 & 28.7 & 0      & 86.77/96.37  & 10.38 \\
Ours, base                  & -     & -     & 24.4 & 0      & 97.94/107.54 & 9.30 \\
\hline
Ours, dt                    & 32.66 & 0.959 & 24.4 & 1038   & 11.17/20.35 & 49.14\\
Ours, dt, diffuse MLP       & 32.73 & 0.960 & 48.8 & 1038   & 14.58/23.77 & 42.07\\
Ours, dt, dift              & 32.58 & 0.960 & 48.8 & 1435  & 11.31/20.60 & 48.54 \\
\hline
Ours, dt, dift, PSH-128     & 32.02 & 0.957 & 30.3 & 1435  & 9.07/18.38 & 54.41\\
Ours, dt, dift, PSH-256     & 32.23 & 0.956 & 53.3 & 1435  & 9.33/18.56 & 53.87\\
Ours, dt, dift, PSH-512     & 32.36 & 0.954 & 139.6 & 1435 & 9.36/18.57 & 53.85\\
\end{tabular}
}
\label{tab:qs-trade-off}
\end{table}

\subsection{Limitation and Discussion} \label{limitation}
We have demonstrated the compelling capability of the NEPHELE cloud platform of distributing the power of neural radiance rendering to everyone to get rid of high-end devices, and realizing a range of unprecedented effects, e.g, building multi-user shareable complex scenes composed of many NeRF objects. However, as a first trial to bring neural radiance rendering to the cloud for multi-user, our NEPHELE system still owns several limitations as follows.

First, from the algorithm side, the proposed i-NOLF relies on a pre-trained density field from i-NGP, so it cannot effectively handle the case where i-NGP cannot reconstruct meaningful geometry, e.g., glossy surfaces. It is promising to address the issue by combining the recent excellent work RefNeRF~\cite{verbin2022refnerf} with i-NGP.
Besides, the empirical strategy employed for hitpoint estimation generates only one point per ray for ultra-fast rendering. However, it cannot work very well for multi-layer objects, e.g., an object composed of a translucent surface in front of a solid surface. In future work, we plan to develop a more comprehensive strategy similar to the translucent material rendering techniques in the traditional rendering pipeline.
Moreover, our current i-NOLF can only reconstruct the object within the lighting environment while capturing it. It is promising and natural to extend the recent surface-based neural relighting technique to the opacity light field in pursuit of more fancy visual effects. 
Finally, although we have deployed our i-NOLF on the cloud, it still requires high-end GPUs for real-time rendering. Further exploration to develop a local renderer similar to~\cite{chen2022mobilenerf} that suits mobile devices is interesting and meaningful.

From the system side, our NEPHELE system only divides and schedules tasks with two categories (i.e., light and heavy), which may introduce external fragmentations to graphic cards. We intend to apply a multi-level hierarchical categorizer to classify tasks to fit compute power of graphic cards. Besides, we plan to extend NEPHELE to be compatible with heterogeneous GPU resources. Moreover, we attempt to construct a RayFarm with high-performance storage systems to speed up access to pre-rendered rays. Collaborated with RayFarm, RenderFarms can eliminate overhead to locate rendered rays and store new ones in the storage pool.

\section{conclusion} \label{sec:conclusion}

We have presented NEPHELE, a neural platform for highly realistic cloud radiance rendering. In stark contrast with existing NR approaches, our NEPHELE allows for more powerful rendering capabilities by combining multiple remote GPUs and facilitates collaboration by allowing multiple people to view the same NeRF scene simultaneously.
For lightweight, real-time neural representation with scalability in our cloud-based scenarios, we introduce i-NOLF to employ opacity light fields, analogous to i-NGP. Our i-NOLF further employs the Lumigraph with geometry proxies, enabling ultra-efficient neural radiance rendering in a one-query-per-ray manner with fast ray querying. We  subsequently adopt a tony MLP for the local opacity lumishperes, achieving high-quality view-dependent rendering. Instead of using muti-resolution hashing in i-NGP, we adopt Perfect Spatial Hashing to significantly improve the cache coherence, with a single-resolution encoding design to further reduce
the cache traffic. 
We also tailor a task scheduler accompanied by our i-NOLF representation, with a ray-level scheduling design to maintain the resiliency of rendering jobs. To this end, we conduct a comprehensive cloud platform with various cooperated modules, i.e., render farms, task assigner, frame composer, and detailed streaming strategies.
Extensive qualitative and quantitive results on various datasets demonstrate the effectiveness of our i-NOLF for cloud-based radiance rendering scenarios. We further showcase various applications of our NEPHELE that are unseen before, ranging from cloud radiance rendering in VR/AR to experiencing, sharing, and assembling various NeRF assets.
With such uniqueness, we believe our approach paves the way to democratize the accessible use of radiance rendering through a novel cloud-based paradigm, enabling numerous potential applications for entertainment and immersive experience in VR/AR.

\bibliographystyle{ACM-Reference-Format}
\bibliography{ref}

\end{document}